\documentclass[prd,superscriptaddress,amsfonts,amssymb,amsmath,natbib,showpacs,nofootinbib,10pt]{revtex4-2}

\usepackage{xcolor}
\usepackage{comment}
\usepackage{graphicx}
\usepackage{dcolumn}
\usepackage{bm}
\usepackage{hyperref}
\usepackage[mathlines]{lineno}
\usepackage{subfigure}
\usepackage{blindtext}
\usepackage{amsmath}
\usepackage{multirow}
\begin{document}

\title{A comprehensive analysis of the compact phase space for Hu-Sawicki $f(R)$ dark energy models including spatial curvature}

\author{Kelly MacDevette}
\email[]{mcdkel004@myuct.ac.za}
\affiliation{Department of Mathematics and Applied Mathematics, University of Cape Town, Rondebosch 7700, Cape Town, South Africa}
\author{Peter Dunsby}
\email[]{peter.dunsby@uct.ac.za}
\affiliation{Department of Mathematics and Applied Mathematics, University of Cape Town, Rondebosch 7700, Cape Town, South Africa}
\author{Saikat Chakraborty}
\email[]{saikatnilch@gmail.com}
\affiliation{Center for Space Research, North-West University, Mahikeng 2745, South Africa}
\date{\today}

\begin{abstract}
We present a comprehensive dynamical systems analysis of homogeneous and isotropic Friedmann-La\^{i}matre-Robertson-Walker  cosmologies in the Hu-Sawicki $f(R)$ dark energy model for the parameter choice $\{n,C_1\}=\{1,1\}$. For a generic $f(R)$ theory, we outline the procedures of compactification of the phase space, which in general is 4-dimensional. We also outline how, given an $f(R)$ model, one can determine the coordinate of the phase space point that corresponds to the present day universe and the equation of a surface in the phase space that represents the $\Lambda$CDM evolution history. Next, we apply these procedures to the Hu-Sawicki model under consideration. We identify some novel features of the phase space of the model such as the existence of invariant submanifolds and 2-dimensional sheets of fixed points. We determine the physically viable region of the phase space, the fixed point corresponding to possible matter dominated epochs and discuss the possibility of a non-singular bounce, re-collapse and cyclic evolution. We also provide a numerical analysis comparing the $\Lambda$CDM evolution and the Hu-Sawicki evolution.
\end{abstract}

\maketitle

\section{Introduction}
A hundred years after it was formulated, Einstein’s theory of General Relativity (GR) remains the best description of the gravitational action, surviving observational tests ranging from the solar system to extragalactic scales. Despite this success, high precision astrophysical and cosmological observations conducted over the past two decades (supernovae Ia, cosmic microwave background anisotropies, large scale structure formation, baryon oscillations and weak lensing measurements) \cite{Ostriker:1995rn,*SupernovaCosmologyProject:1998vns,*SupernovaSearchTeam:1998fmf,*SupernovaSearchTeam:2003cyd,*SupernovaCosmologyProject:2003dcn,*SupernovaSearchTeam:2004lze,*SNLS:2005qlf,*WMAP:2003elm,*WMAP:2006bqn,*SDSS:2003eyi,*SDSS:2004kqt,*2dFGRS:2005yhx,*SDSS:2005xqv,*Blake:2005jd,*Jain:2003tba} seems to suggest that, at least on cosmological scales, General Relativity might not represent a complete description of gravity. In particular, the late-time acceleration of the Universe cannot be described within the framework of General Relativity without introducing additional exotic matter degrees of freedom - known as Dark Energy. Although inferred by fitting observational data to our preferred geometry of the universe (the Robertson-Walker metric), Dark Energy has not been directly observed. Currently the Concordance (or $\Lambda$CDM) model is regarded as the theoretical framework that best fits all available data. Here a Cosmological Constant dominates the present day universe, driving late-time acceleration, while ordinary matter is dominated by Cold Dark Matter (CDM), which form the potential wells for large-scale structure formation. Despite its success, the $\Lambda$CDM model is plagued by significant fine-tuning problems related to the vacuum energy scale, so it is important to consider other theoretical alternatives, which are able to describe the main features of the universe, without the introduction of a cosmological constant or dynamical Dark Energy.

One of the most popular alternatives to the Concordance Model is based on modifications of standard Einstein gravity \cite{Sotiriou:2008rp,DeFelice:2010aj}. Such models became popular in the 1980’s because it was shown that they naturally admit a phase of accelerated expansion which could be associated with inflation in the early universe. This property has led to the idea that late-time acceleration could also have a geometrical origin and that there is a connection between Dark Energy and a non-standard behaviour of gravitation on cosmological scales. 

Among one of the most widely studied of these geometrical dark energy models is the so called $f(R)$ theories of gravity \cite{Capozziello:2002rd}. These models are based on gravitational actions which are non-linear in the Ricci curvature R. Here the higher order curvature corrections are written as an energy-momentum tensor of geometrical origin describing an “effective” source term on the right hand side of the standard Einstein field equations. This leads naturally to a geometrical origin of the late-time acceleration of the Universe. $f(R)$ gravity is characterised by the existence of a propagating scalar degree of freedom (called scalaron or curvaton) $\varphi= f'(R)$ as apparent from the trace field equation
\begin{equation}
    Rf'(R)-2f(R)+3\Box f'(R) = T,
\end{equation}
$T$ being the trace of the energy momentum tensor \cite{Sotiriou:2008rp,DeFelice:2010aj}. GR is the trivial case of $f(R)$ for which $f'(R)=1$ and the scalar degree of freedom vanishes. One of the well-known late-time $f(R)$ models is the so-called Hu-Sawicki model \cite{Hu:2007nk}, which, in its most generic form, is a 3-parameter $f(R)$ model. We will be particularly focusing on this important model in this paper.

Because of the fourth order nature of these theories, the cosmological dynamics is considerably richer (and more complicated) than their GR counterparts, making the theory of dynamical systems (DS) ideally suited for studying them. In fact, describing these cosmologies using the dynamical systems approach provides a relatively simple method for finding exact solutions (via their corresponding fixed points) and obtaining a (qualitative) description of the global dynamics of these models \cite{Abdelwahab:2007jp,Carloni:2004kp,Carloni:2007br,Carloni:2015jla,Abdelwahab:2011dk,Goheer:2007wu,Goheer:2007wx}. This approach has been very successful in shedding light on the dynamics of a range of homogeneous cosmologies including the Friedmann-La\^{i}matre-Robertson-Walker (FLRW), Bianchi I \& 5 \cite{Leach:2006br,Goheer:2007wu} and Kantowski-Sachs models \cite{Solomons:2001ef}. It has also proved useful in identifying those orbits in phase-space (expansion histories) which are close to the $\Lambda$CDM model, which can then be used as suitable background cosmologies for the study of the growth of large scale structure \cite{Ananda:2008gs,Ananda:2008tx,Carloni:2007yv,Abebe:2013zua}.

The paper is organised as follows: In Sec.\ref{sec:gen} we consider a generic $f(R)$ theory and review the DS approach to studying these models. In particular, we describe how to compactify the phase-space and identify surfaces containing orbits which represent $\Lambda$CDM evolution, as well as the point in phase-space corresponding to the present universe (based on the current values of the Hubble and deceleration parameters). In \ref{sec:HS} we apply this framework to the Hu-Sawicki $f(R)$ dark energy model for the parameter choice $\{n,C_1\}=\{1,1\}$, obtaining a complete description of the dynamics of these models including spatial curvature. We determine the physically viable region of the phase space, the fixed point corresponding to possible matter dominated epochs and discuss the possibility of a non-singular bounce, re-collapse and cyclic evolution. We also provide a numerical analysis comparing the $\Lambda$CDM evolution and the Hu-Sawicki evolution, setting initial conditions so that the HS model is dynamically indistinguishable from the Concordance model. Throughout the paper we use the metric signature $(-,+,+,+)$ and the convention $\kappa=8\pi G=1$.

\section{Dynamical systems approach for generic $f(R)$ cosmology}
\label{sec:gen}
We start by writing the field equation for $f(R)$ gravity in the presence of a perfect fluid in a homogeneous and isotropic FLRW background \cite{Sotiriou:2008rp,DeFelice:2010aj}
\begin{subequations}\label{fe}
\begin{eqnarray}
&& 3f'\left(H^{2}+\frac{k}{a^2}\right) = \rho_{\rm eff} \equiv \rho+\rho_{R},\label{fe1}\\
&& -f'\left(2\dot{H}+3H^{2}+\frac{k}{a^2}\right) = P_{\rm eff} \equiv P+P_{R},\label{fe2}
\end{eqnarray}
\end{subequations}
where the prime denotes a derivative with respect to $R$ and we have defined the curvaton energy density and pressure as
\begin{subequations}
\label{curv}
\begin{eqnarray}
&& \rho_{R} \equiv \frac{1}{2}(Rf'-f)-3H\dot{f'},\\
&& P_{R} \equiv \ddot{f'}+2H\dot{f'}-\frac{1}{2}(Rf'-f).
\end{eqnarray}
\end{subequations}
The curvaton equation of state parameter is
\begin{equation}\label{w_R}
    w_R \equiv \frac{P_R}{\rho_R} = \frac{\ddot{f'}+2H\dot{f'}-\frac{1}{2}(Rf'-f)}{\frac{1}{2}(Rf'-f)-3H\dot{f'}},
\end{equation}
and the effective equation of state parameter of the universe is 
\begin{equation}\label{eq:w_eff}
    w_{\rm eff} \equiv \frac{P_{\rm eff}}{\rho_{\rm eff}} = \frac{P+P_R}{\rho+\rho_R} = -\frac{2\dot{H}+3H^{2}+k/a^2}{3\left(H^2 + k/a^2\right)}.
\end{equation}
If the perfect fluid is a barotropic one with an equation of state parameter $\frac{P}{\rho}=w$, then $w,\,w_R$ and $w_{\rm eff}$ are related as
\begin{equation}\label{www}
    w_{\rm eff} = w\frac{\rho}{\rho_{\rm eff}} + w_R\frac{\rho_R}{\rho_{\rm eff}}.
\end{equation}
Since matter is minimally coupled to curvature, the perfect fluid and the curvaton evolve independently according to their own continuity equations \cite{Capozziello:2008qc}
\begin{subequations}
\begin{eqnarray}
&& \dot{\rho} + 3H(1+w)\rho = 0,
    \\
&& \dot{\rho}_{R} + \left(3H(1+w_{R})+\frac{\dot{R}f''}{f'^2}\right)\rho_{R} = - \left(2H-3\frac{\dot{R}f''}{f'}\right)\frac{k}{a^2} + 3\frac{H^2\dot{R}f''}{f'}.
\end{eqnarray}
\end{subequations}
The matter continuity equation gives the usual scaling
\begin{eqnarray}
    \rho = 3H_0^{2}\Omega_{0}a^{-3(1+w)},
\end{eqnarray}
where we have used the definition of the observable quantity matter density parameter $\Omega_{0}=\frac{\rho_{0}}{3H_0^2}$, with the subscript `$0$' being used to denote the present day values.

It is worth mentioning here two important conditions for physical viability of any $f(R)$ gravity model:
\begin{itemize}
    \item $f'(R)<0$ makes the scalar degree of freedom appearing in the theory a ghost. To eradicate the possibility of a ghost degree of freedom, one must require $f'(R)>0$ for all $R$.
    \item $f''(R)<0$ is related to unstable growth of curvature perturbation in the weak gravity limit (this is also known as the Dolgov-Kawasaki instability, as this was first suggested by Dolgov and Kawasaki in \cite{Dolgov:2003px}). Therefore one requires $f''(R)>0$ at least during the early epoch of matter domination.
\end{itemize}
\subsection{Phase space analysis for $f(R)$ gravity}
In this section we present a dynamical system formulation for FLRW cosmologies in $f(R)$ gravity following the Refs.\cite{Carloni:2004kp,Carloni:2007br}. We start by writing the Friedmann equation \eqref{fe1} as follows
\begin{eqnarray}\label{fried}
    H^2 = \frac{1}{3}\frac{\rho}{f'} + \frac{1}{6f'}(Rf'-f) - H\frac{\dot{f}'}{f'} - \frac{k}{a^2}.
\end{eqnarray}
Next, we divide both side by $3H^2$ and define a set of Hubble-normalized dimensionless dynamical variables 
\begin{equation}\label{var_def}
     \tilde{x}=\frac{\dot{f}'}{f'}\frac{1}{H}, \qquad
      \tilde{v}=\frac{1}{6}\frac{R}{H^2}, \qquad \tilde{y}=\frac{1}{6}\frac{f}{f'}\frac{1}{H^2},\qquad
     \tilde{\Omega}=\frac{1}{3}\frac{\rho}{f'}\frac{1}{H^2}, \qquad
      \tilde{K}=\frac{k}{a^2}\frac{1}{H^2},
\end{equation}
that are related by the constraint equation
\begin{eqnarray}\label{constr_1}
    \tilde{\Omega} + \tilde{v} - \tilde{y} - \tilde{x} - \tilde{K} = 1.
\end{eqnarray}
Differentiating the dynamical variables with respect to the e-folding number $N\equiv\ln a$, one obtains the following autonomous system of equations
\begin{subequations}\label{dynsys_1}
\begin{eqnarray}
&& \frac{d\tilde{x}}{dN} = -2\tilde{v}+4+3\tilde{x}-3(1+w)\tilde{\Omega}+4\tilde{K}-\tilde{x}^2-\tilde{x}\tilde{v}+\tilde{x}\tilde{K}, \\
&& \frac{d\tilde{v}}{dN} = \tilde{v}\left(\Gamma \tilde{x}-2\tilde{v}+2\tilde{K}+4\right),\\
&& \frac{d\tilde{K}}{dN} = -2\tilde{K}\left(\tilde{v}-\tilde{K}-1\right),\\
&& \frac{d\tilde{\Omega}}{dN}
    = -\tilde{\Omega}\left(-1+3w+\tilde{x}+2\tilde{v}-2\tilde{K}\right),\\
&& \frac{d\tilde{y}}{dN} = \Gamma \tilde{x}\tilde{v} + \tilde{y}(2\tilde{K}-2\tilde{v}-\tilde{x}+4),
\end{eqnarray}
\end{subequations}
where we had defined an auxiliary quantity
\begin{equation}
    \Gamma \equiv \frac{f'}{Rf''},
\end{equation}
that is explicitly dependent on the functional form of $f(R)$. The system of dynamical equations \eqref{dynsys_1}, along with the Friedmann constraint \eqref{constr_1}, constitutes the dynamical system for FLRW cosmologies in $f(R)$ gravity. Because of the existence of the constraint, one of the dynamical variables is redundant, and the resulting phase space is actually 4-dimensional. 

In order to close the system, $\Gamma$ must be expressed as a function of the dynamical variables. This can be done by noting that the relation
\begin{equation}\label{invert}
    \frac{\tilde{v}}{\tilde{y}} = \frac{Rf'}{f}
\end{equation}
can, in principle, be inverted to obtain $R=R\left(\frac{\tilde{v}}{\tilde{y}}\right)$. Using this one can obtain $\Gamma=\Gamma\left(\frac{\tilde{v}}{\tilde{y}}\right)$. As mentioned in \cite{Carloni:2007br}, the success of this particular dynamical system formulation crucially depends on the invertibility of the relation \eqref{invert}. This is a limitation of this formulation. For example, the most generic case of the Hu-Sawicki $f(R)$ model \cite{Hu:2007nk} cannot be treated with this approach and a different formulation is necessary. A possible solution to this issue was devised in \cite{Carloni:2015jla}. In this paper we will focus on the only particular case of the Hu-Sawicki $f(R)$ that can in fact be treated with this formulation \cite{Kandhai:2015pyr} 
\footnote{For a historical account of different dynamical system formulations for FLRW cosmology in $f(R)$ gravity, see \cite{Chakraborty:2021mcf}.}.

The Hubble-normalized dynamical variables defined in Eq.\eqref{var_def} are non-compact, i.e., their domain is infinite or semi-infinite. In particular, by definition, all of them diverge at a cosmological bounce or recollapse, that necessarily requires the condition $H=0$. Therefore, there can be interesting physical properties at the infinity of the phase space that one misses using this formulation. Also, one can justifiably take the e-folding $N\equiv\ln a$ as a phase space time variable as long as one confines the attention to an expanding universe only, as $\dot{N}=H$ is always positive. For a contracting universe one should alter the definition as $N\equiv-\ln a$ and in general one should define the phase space time variable as $N\equiv\frac{|H|}{H}\ln a$. Clearly this definition is discontinuous at a cosmological bounce or recollapse. As we will see in the next subsection, this shortcoming is not present in the compact phase space formulation.
\subsection{Compactifying the phase space} \label{sec:compact}
In this section we outline a prescription for the compactification of the phase space. We follow the approach of Refs.\cite{Abdelwahab:2011dk,Kandhai:2015pyr}, but generalize it to the case of non-negative global spatial curvature \footnote{For a different compactification approach via the Poincare compactification, see \cite{Abdelwahab:2007jp}}. We start by writing the Friedmann equation \eqref{fe1} as follows
\begin{equation}\label{fried_compact}
 D^2 \equiv  \left(3H + \frac{3}{2}\frac{\dot{f'}}{f'}\right)^{2} + \frac{3}{2}\left(\frac{f}{f'}+\frac{6k}{a^2}\right) = \frac{3\rho}{f'} + \frac{3}{2}R + \frac{9}{4}\left(\frac{\dot{f'}}{f'}\right)^2 .
\end{equation}
Next, we divide both side by $D^2$ and define a set of \emph{compact} dynamical variables 
\begin{equation}\label{var_def_compact}
 x=\frac{3}{2}\frac{\dot{f}^{\prime}}{f^{\prime}}\frac{1}{D}, \qquad v=\frac{3}{2}\frac{R}{D^{2}}, \qquad y=\frac{3}{2}\frac{f}{f^{\prime}}\frac{1}{D^{2}}, \qquad
 \Omega=\frac{3\rho}{f^{\prime}}\frac{1}{D^{2}}, \qquad Q=\frac{3H}{D}, \qquad K=\frac{9k}{a^2}\frac{1}{D^2},
\end{equation}
which are now related by two constrained equations
\begin{subequations}
\begin{eqnarray}
&& (Q+x)^{2}+K+y=1 \label{eq:compconstraint},\\
&& \Omega+x^{2}+v=1 \label{eq:friedconstraint},
\end{eqnarray}
\end{subequations}
that comes directly from Eq.\eqref{fried_compact}. The dynamical variables defined in Eq.\eqref{var_def_compact} are compact because, requiring $f'(R)>0$, restricting ourselves to non-negative global spatial curvature and keeping in mind the two constraints \eqref{eq:compconstraint},\eqref{eq:friedconstraint}, one gets the following finite domains for the dynamical variables 
\begin{equation}
\begin{aligned}
& -1 \leq x \leq 1, \quad
   0 \leq \Omega \leq 1, \quad
  -2 \leq Q \leq 2, 
  \\
& 0 \leq v \leq 1, \quad
  0 \leq y \leq 1, \quad
  0 \leq K \leq 1.
\end{aligned}
\end{equation}
It is straightforward to find a relationship between the compact dynamical variables and their non-compact counterparts
\begin{equation}\label{var_reln}
     \tilde{x}=2\frac{x}{Q}, \qquad 
      \tilde{v}=\frac{v}{Q^2}, \qquad
      \tilde{y}=\frac{y}{Q^2},\qquad
     \tilde{K}=\frac{K}{Q^2}, \qquad
      \tilde{\Omega}=\frac{\Omega}{Q^2}.
\end{equation}

To write an autonomous dynamical system, we need to define a new phase space time variable $\tau$ such that
\begin{equation}
    d\tau \equiv Ddt,
\end{equation}
Differentiating the compact dynamical variables with respect to $\tau$ and using the field equations results in a system of six first order autonomous differential equations. Because of the existence of two constraint equations \eqref{eq:compconstraint} and \eqref{eq:friedconstraint}, two of the dynamical variables are redundant. Therefore, the compact phase space, just like it's non-compact counterpart, is actually 4-dimensional. We choose to eliminate $y$ and $\Omega$ and write the dynamical system as follows 
\begin{widetext}
    \begin{subequations}
    \begin{eqnarray}
       && \frac{dv}{d\tau} = -\frac{1}{3}v\left((Q+x)\left(2v - (1+3w)(1-x^{2}-v) + 4xQ\right) - 2Q - 4x + 2x\Gamma(v-1) + 4xK\right),
       \\
       && \frac{dx}{d\tau} = \frac{1}{6}\Bigg[-2x^{2}v\Gamma + (1-3w)(1-x^{2}-v) + 2v + 4\left(x^{2}-1\right)\left(1-Q^{2}-xQ\right) + x(Q+x)\left((1+3 w)(1-x^{2}-v) - 2v\right)\nonumber\\
       && \qquad + 4K(1-x^2)\Bigg],
       \\
       && \frac{dQ}{d\tau} = \frac{1}{6}\Bigg[-4xQ^{3} + xQ(5+3w)(1-xQ) - Q^{2}(1-3w) - Qx^{3}(1+3w) - 3vQ(1+w)(Q+x)\nonumber\\
       && \qquad + 2v(1-Qx\Gamma) - 2K(1+2xQ)\Bigg],
       \\
       && \frac{dK}{d\tau} = -\frac{1}{3}K\left((Q+x)(-(1+3w)(1-x^{2}-v) + 4xQ + 2v) + 4x(K-1) + 2xv\Gamma\right).
    \end{eqnarray}
    \label{eq:compactdynsys}
    \end{subequations}
\end{widetext}
The system can be closed by inverting the relation
\begin{equation}\label{invert_compact}
    \frac{v}{y} = \frac{Rf'}{f},
\end{equation}
to find $R=R\left(\frac{v}{y}\right)$ and using it to write $\Gamma=\Gamma\left(\frac{v}{y}\right)$, provided, of course, that the relation \eqref{invert_compact} is invertible. Note that, unlike in the non-compact phase space analysis, the definition of the phase space time variable $\tau$ is \emph{not} discontinuous at a cosmological bounce or recollapse, as $\dot{\tau}=D$ is always positive definite by definition.

A compact phase space analysis helps analyze the global features of the solution space of a theory, some of which might not be apparent from a non-compact phase space analysis. From this point onwards, we will use only the compact phase space analysis for all of our subsequent analysis.
\subsection{$\Lambda$CDM point}\label{subsec:lcdm_pt_gen}
In this section we will try to determine which point in the 4-dimensional phase space $v$-$x$-$Q$-$K$ corresponds to the present day universe. In other words, we will outline how to calculate the values of the compact dynamical variables $v,\,x,\,Q,\,K$ based on the observed values of different cosmological parameters. To this aim, let us first introduce two sets of dimensionless parameters. The first set is the usual matter density parameter ($\Omega_m$) and spatial curvature density parameter ($\Omega_k$)
\begin{equation}
    \Omega_m\equiv\frac{\rho}{3H^2}, \qquad \Omega_{k}\equiv\frac{k}{a^{2}H^{2}},
\end{equation}
where we have deliberately used the subscript `$m$' in $\Omega_m$ to make it apparent that it is different from the compact dynamical variable $\Omega$ defined in Eq.\eqref{var_def_compact} ($\Omega\rightarrow\Omega_m$ only at the GR limit $f'\rightarrow1$). The second set is the dimensionless cosmographic parameters, namely the dimensionless Hubble parameter ($h$), deceleration parameter ($q$), jerk parameter ($j$) and snap parameter ($s$) parameter \cite{Visser:2003vq,Dunajski:2008tg}
\begin{equation}
    h\equiv\frac{H}{H_0}, \qquad
    q\equiv-\frac{\ddot{a}}{aH^2}, \qquad
    j\equiv\frac{\dddot{a}}{aH^3}, \qquad
    s\equiv\frac{a^{(4)}}{aH^4},
\end{equation}
where $H_0$ is the present day value of the Hubble parameter. Both the matter density parameters and the cosmographic parameters are sets of observable quantities. 

Assuming that the dominant hydrodynamic fluid content in the universe is given by the cold dark matter (CDM) that is modelled by a dust fluid, the Hubble parameter and its first derivative are given by
\begin{subequations}
\begin{eqnarray}
    && H^2 = \frac{\rho}{3f'} + \frac{1}{6f'}(Rf'-f) - H\frac{\dot{f}'}{f'} - \frac{k}{a^2},
    \label{H}
    \\
    && \dot{H} = - \frac{1}{2f'}\left[\frac{1}{3}\rho + \ddot{R}f'' + \dot{R}^{2}f''' - Hf''\dot{R}\right] + \frac{3}{2}\frac{k}{a^2},
    \label{Hdot}
\end{eqnarray}
\end{subequations}
It can be verified that Eqs.(\eqref{H},\eqref{Hdot}) are completely equivalent to Eqs.(\eqref{fe1},\eqref{fe2}). The Ricci scalar and its first two derivatives are as follows
\begin{subequations}
  \begin{eqnarray}
    && R = 6\left(\dot{H} + 2H^2 + \frac{k}{a^2}\right),
   \label{eq:ricciscalar} \\
    && \dot{R} = 6\left(\ddot{H} + 4H\dot{H} - 2H\frac{k}{a^2}\right),
    \\
    && \ddot{R} = 6\left[\dddot{H} + 4H\ddot{H} + 4\dot{H}^2 + \left(4H^{2}-2\dot{H}\right)\frac{k}{a^2}\right].
  \end{eqnarray}
\end{subequations}
Using the matter abundance parameters and the cosmographic parameters, the Ricci scalar and its first two derivatives can be expressed as purely algebraic combinations of them \cite{Capozziello:2008qc}
\begin{subequations}\label{R_Rdots}
  \begin{eqnarray}
    && \frac{R}{H_0^2} = 6h^{2}(1-q+\Omega_k),
    \\
    && \frac{\dot{R}}{H_0^3} = 6h^{3}(j-q-2-2\Omega_k),
    \\
    && \frac{\ddot{R}}{H_0^4} = 6h^{4}\left(s+q^{2}+8q+6+2(3+q)\Omega_k\right).
\end{eqnarray}
\end{subequations}
Eqs.(\eqref{H},\eqref{Hdot}) can also be expressed using the cosmographic parameters as purely algebraic equations i.e. not explicitly containing any time derivatives
\begin{subequations}\label{eq:initcalc}
\begin{eqnarray}
&& \mathcal{A}f' +\frac{f}{H^2}+\mathcal{B}\left(H^2f''\right)-6\Omega_m=0,\label{eq:initcalc1}
\\
&& \mathcal{C}f'-\mathcal{D}\left(H^2f''\right)-\mathcal{G}\left(H^4f'''\right)-\Omega_m=0,\label{eq:initcalc2}
\end{eqnarray}
\end{subequations}
with the coefficients $\mathcal{A},\,\mathcal{B},\,\mathcal{C},\,\mathcal{D},\,\mathcal{G}$ being different algebraic combinations of the cosmographic parameters and the spatial curvature density parameter
\begin{subequations}\label{bigeq}
\begin{eqnarray}
&& \mathcal{A} = 6q,\\
&& \mathcal{B} = 36(j-q-2-2\Omega_{k}),\\
&& \mathcal{C} = 3\Omega_k+2+2q,\\
&& \mathcal{D} = 6( 8 -j +9q +q^2 +2\Omega_k(4+q) +s),\\
&& \mathcal{G} = 36(j-q-2-2\Omega_k)^2.
\end{eqnarray}
\end{subequations}

Henceforth we use the subscript `$0$' to denote the parameter values in the current epoch. Since the value of the scale factor itself bears no dynamical significance, it is conventional to take $a_0=1$. Also, by definition, the present day value of the dimensionless Hubble parameter is $h_0=1$. Other cosmographic parameters are quantities that can be calculated from the observational data once a parametrization for the dark energy equation of state is fixed. Using the Chevallier-Polarski-Linder (CPL) parameterization \cite{CP2001,L2003} for the dark energy equation of state
\begin{equation}
    w_{\rm DE} = w_0 + w_a(1-a)
\end{equation}
and fixing the free parameters to those corresponding to the $\Lambda$CDM model $(w_0,w_a)=(-1,0)$ results in the following expressions for $(q_0,j_0,s_0)$
\begin{subequations}
\begin{eqnarray}
&& q_0 = \frac{1}{2}-\frac{3}{2}\Omega_{\Lambda0},\\
&& j_0 = \Omega_k + 1,\\
&& s_0 = 1-\frac{9}{2}\Omega_{m0},
\end{eqnarray}
\end{subequations}
where $\Omega_{\Lambda}$ is used to denote the dark energy density parameter. The values of different density parameters, taken from Planck 2018 data \cite{Planck2020}, are
\begin{equation}\label{dens_param_0}
    (\Omega_{\Lambda0},\Omega_{k0},\Omega_{m0})=(0.6843,0.0007,0.315).
\end{equation}
This gives the present day values for the cosmographic parameters within the $\Lambda$CDM model as
\begin{equation}\label{cosm_param_0}
    (q_0,j_0,s_0)=(-0.52645,1.0007,-0.4175).
\end{equation}
The values in Eqs.(\eqref{dens_param_0},\eqref{cosm_param_0}) allow us to calculate the present day values of the Ricci scalar and its derivatives using Eq.\eqref{R_Rdots}
\begin{equation}\label{R_Rdots_0}
    \left(\frac{R_0}{H_0^2},\frac{\dot{R}_0}{H_0^3},\frac{\ddot{R}_0}{H_0^4}\right)=(9.1629,-2.8455,9.9091),
\end{equation}
as well as the present day values of the coefficients in \eqref{bigeq}
\begin{equation}
 (\mathcal{A}_0,\mathcal{B}_0,\mathcal{C}_0,\mathcal{D}_0,\mathcal{G}_0) 
 =(-3.1587,-17.0730,0.9492,12.7546,8.0969).
\end{equation}
With the known values of these coefficients, one can see now that Eq.\eqref{eq:initcalc} becomes a pair of algebraic equations with constant coefficients for $\frac{f_0}{H_0^2},f'_0,H_0^{2}f''_0,H_0^{4}f'''_0$. If one now specifies an explicit functional form of $f(R)$, the two equations \eqref{eq:initcalc1} and \eqref{eq:initcalc2} can be used to determine the value of up to two a-priori unspecified model parameter of the theory (modulo a factor of $H_0$ that is used to make the model parameters dimensionless). 

Note that if there are more than two unspecified model parameters, one requires more equations additional to \eqref{eq:initcalc1},\eqref{eq:initcalc2}, which are obtained by taking time derivatives of Eq.\eqref{Hdot} and considering cosmographic parameters of order higher than the snap parameter. On the other hand, if there is only one unspecified model parameter, then the two equations \eqref{eq:initcalc1},\eqref{eq:initcalc2} are redundant to find its value. In this situation, one should consider the reduced set of cosmographic parameters $\{q,j\}$ and use their present values to determine the value of the model parameter. Eq.\eqref{eq:initcalc1} now suffices for this purpose. Eq.\eqref{eq:initcalc2} in this case does not provide an additional equation to determine the model parameter, as it contains $s$ whose present day value will not be required anymore. 

Once the value(s) of the a-priori unspecified model parameter(s) is determined (modulo the factor of $H_0$), one can calculate the values of the compact dynamical variables corresponding to the present day observable universe using the values given in Eqs.(\eqref{cosm_param_0},\eqref{R_Rdots_0}). Henceforth we will call this point in the phase space the $\Lambda$CDM point. In Sec.\ref{subsec:lcdm_hs} we will calculate the phase space coordinates of the $\Lambda$CDM point for the specific case of the Hu-Sawicki $f(R)$ model that we consider in this paper.
\subsection{$\Lambda$CDM surface}
The cosmological evolution in the $\Lambda$CDM model can be written as
\begin{eqnarray}
h(a) &=& \sqrt{\Omega_{m0}a^{-3} + \Omega_{\Lambda0} - \Omega_{k0}a^{-2}} \nonumber\\
    &=& \sqrt{\Omega_{m0}a^{-3} + \Omega_{\Lambda0} - (\Omega_{m0}+\Omega_{\Lambda0}-1)a^{-2}},
    \label{lcdm}
\end{eqnarray}
where $\Omega_{m0},\,\Omega_{\Lambda0},\,\Omega_{k0}$ has the values specified in \eqref{dens_param_0} and the second line comes from the first line by writing the equation at the present epoch. The $\Lambda$CDM evolution history can be specified by the necessary and sufficient condition\footnote{By necessary and sufficient condition, we imply that Eq.\eqref{lcdm}, differentiated twice, gives Eq.\eqref{lcdm_cosm} and Eq.\eqref{lcdm_cosm}, integrated twice, gives back Eq.\eqref{lcdm}.} \cite{Dunajski:2008tg}
\begin{equation}\label{lcdm_cosm}
    j = \Omega_k + 1.
\end{equation}
This is a purely kinematic condition specifying \emph{how} the universe evolves, with no information on what is the inherent dynamical model for the evolution. There can be models other than the $\Lambda$CDM that obey the same cosmographic requirement \eqref{lcdm_cosm} and therefore produces a cosmological evolution identical to \eqref{lcdm}. Such models are kinematically degenerate with $\Lambda$CDM at the background level. For an $f(R)$ model to exactly mimic the $\Lambda$CDM evolution history at the background level, it must obey the condition \eqref{lcdm_cosm}\footnote{For reconstruction of $f(R)$ models based on the $\Lambda$CDM evolution history \eqref{lcdm}, see \cite{Dunsby:2010wg}. For a generic dynamical system analysis of $\Lambda$CDM mimicking $f(R)$ models, see \cite{Chakraborty:2021jku}.}

The deceleration parameter $q$ can be expressed in terms of the dynamical variables as
\begin{equation}
    q = 1+\tilde{K}-\tilde{v} = 1+\frac{K-v}{Q^2}.
\end{equation}
The jerk parameter $j$ can be expressed in terms of the deceleration parameter and it's time derivative.
\begin{equation}
    j = 2q^{2} + q - \frac{1}{H}\frac{dq}{dt} = 2q^{2} + q - \frac{3}{Q}\frac{dq}{d\tau}.
\end{equation}
Using the dynamical equations \eqref{eq:compactdynsys}, $j$ can be expressed as a function of the dynamical variables
\begin{equation}
   j=j(K,x,v,Q),
\end{equation}
whose explicit functional form we do not write here as the relevant calculations can be tackled numerically. Eq.\eqref{lcdm_cosm} now provides an algebraic constraint which specifies a 3-dimensional hypersurface over the entire 4-dimensional phase space $K$-$x$-$v$-$Q$. Henceforth we will call it the $\Lambda$CDM surface. If an $f(R)$ theory is to produce a cosmic evolution history that is asymptotically $\Lambda$CDM, then there must be a De-Sitter saddle or stable fixed point lying on this hypersurface and a phase trajectory that asymptotically approaches this point. If an $f(R)$ theory is to produce a cosmic evolution history that is indistinguishable from the $\Lambda$CDM evolution \eqref{lcdm} at all times, then there must be an unstable or a saddle fixed point corresponding to a matter dominated phase on this surface, a De-Sitter saddle or stable fixed point on this surface, and a phase trajectory connecting these two fixed points it must produce a phase trajectory lying entirely on this surface connecting the two fixed points.

\section{Dynamical systems analysis applied to the Hu-Sawicki $f(R)$ Model}
\label{sec:HS}

A theoretically viable $f(R)$ alternative to dark energy must satisfy the following criteria
\begin{itemize}
    \item $f(R) \to R$ for $R \to 0$.
    \item $f(R) \to R-2\Lambda$ ($\Lambda>0$) for $R \to \infty$.
    \item $f'(R)>0$ for all $R$ so that the curvature degree of freedom is always physical.
    \item $f''(R)>0$ for all $R$ so that at no point of time there is an unbounded growth of curvature perturbation. 
\end{itemize}
The Hu-Sawicki $f(R)$ model \cite{Hu:2007nk} was specifically designed so as to satisfy all of these conditions. In its most generic form, the Hu-Sawicki (henceforth HS) model is a 3-parameter theory which has the following form 
\begin{equation}\label{HS}
    f(R) = R - \frac{C_1 R^n}{C_2 R^n + 1}.
\end{equation}
Taking the limit $R\to\infty$, one can read off $\Lambda=\frac{C_1}{C_2}$, so that $\frac{C_1}{C_2}$ must be positive.
As was pointed out in Ref.\cite{Kandhai:2015pyr}, only the particular case $n=1=C_1$ can be studied with the present dynamical system formulation since only for this special case the relation \eqref{invert_compact} is invertible. Henceforth we will confine our study to this special case only. Choosing to work with the particular case $n=1$ does not, in any way, make our subsequent analysis any less relevant, because it was shown in Ref.\cite{Santos} that $n$ remains completely unconstrained by current cosmological data. Since we live in a universe with a positive value of the cosmological constant, choosing $C_1=1$ means now that $C_2$ must have a positive value.
\subsection{Compact dynamical system}
\label{subsec:compact_HS}
For $n=1=C_1$, the relation \eqref{invert_compact} is invertible and one can find 
\begin{eqnarray}
    \Gamma = \frac{1}{2}\frac{vy}{(v-y)^2}.
\end{eqnarray}
With the above expression for $\Gamma$, the dynamical system \eqref{eq:compactdynsys} becomes singular with a pole of second order on the 3-dimensional hypersurface given by 
\begin{eqnarray}\label{z}
    z\equiv(v-y)=v+(Q+x)^{2}+K-1=0,
\end{eqnarray}
which means one needs to be careful to construct a dynamical system here so as to be able to apply the dynamical system technique. To this goal, we redefine the phase space time variable as
\begin{eqnarray}\label{time_redef}
    d\eta = \frac{d\tau}{\big(v+(Q+x)^{2}+K-1\big)^2},
\end{eqnarray}
so that with respect to this redefined time variable, the new dynamical system becomes regular everywhere. The explicit form of the dynamical system in the compact phase space in presence of a generic perfect fluid and for the HS model with $n=1=C_1$ is as follows
\begin{widetext}
\begin{subequations}\label{eq:compactdynsysHS}
\begin{eqnarray}
&& \frac{dv}{d\eta} = -\frac{1}{3}vz^{2}\Big((Q+x)\big(2v-(1+3w)(1-x^{2}-v)+4xQ\big)-2Q-4x+4xK\Big) - \frac{1}{3}v^{2}x(v-z)(v-1),
\\
&& \frac{dx}{d\eta} = \frac{1}{6}z^{2}\Bigg[(1-3w)(1-x^{2}-v)+2v+4(x^{2}-1)(1-Q^{2}-xQ)+x(Q+x)\big((1+3w)(1-x^{2}-v)-2v\big)\nonumber\\
&& +4K(1-x^2)\Bigg] - \frac{1}{6}x^{2}v^{2}(v-z),
\\
&& \frac{dQ}{d\eta} = \frac{1}{6}z^{2}\Bigg[-4xQ^{3}+xQ(5+3w)(1-xQ)-Q^{2}(1-3w)-Qx^{3}(1+3w)-3vQ(1+w)(Q+x)\nonumber\\
&& +2v-2K(1+2xQ)\Bigg] - \frac{1}{6}v^{2}(v-z)Qx,
\\
&& \frac{dK}{d\eta} = -\frac{1}{3}z^{2}K((Q+x)(-(1+3w)(1-x^{2}-v)+4xQ+2v)+4x(K-1)) - \frac{1}{3}Kxv^{2}(v-z).
\end{eqnarray}
\end{subequations}
\end{widetext}
The variable $z$ from Eq.\eqref{z} follows the dynamical equation
 \begin{eqnarray}
\frac{dz}{d\eta}=&&\frac{1}{3} z\Bigg[v^2 x \Big(k+(Q+x)^2+v-1\Big)-z \bigg(x \left(4 k^2+k \left(8 Q^2+7 v-7\right)+4 Q^4+Q^2 (v-1) (9 w+13)+3 (v-1)^2 (w+1)\right)\nonumber\\
&&+Q \left(3 k (v-1)+3 Q^2 (v-1) (w+1)+3 v^2+3 (v-1)^2 w-4 v+3\right)+x^3 \Big(5 k+3 Q^2 (3 w+5)+2 (v-1) (3 w+2)\Big)\nonumber\\
&&+Q x^2 \left(13 k+Q^2 (3 w+13)+2 (v-1) (6 w+7)\right)+Q (9 w+7) x^4+(3 w+1) x^5\bigg)+ x \Big(k+(Q+x)^2-z-1\Big)^3\Bigg].\label{dyn_z}
 \end{eqnarray}

Interestingly, the singular submanifold $z=0$ now becomes an invariant submanifold in the phase space, by virtue of the time redefinition \eqref{time_redef}. 

The fixed points of the system for the case when the perfect fluid is CDM ($w=0$) are listed in Tab.\ref{tab:fixed_ptsHS}, along with their nature of stability. 
\begin{table*}
	\centering
	\begin{tabular}{|c|c|c|c|c|c|}
		\hline
		Point & $(K,x,v,Q,\Omega, y)$ & Stability & Deceleration parameter & $w_{\rm eff}$ & Cosmology \\
		\hline
		$\mathcal{S}_{1\pm}$ & $\left(K,0,v,\pm\sqrt{1-K-v},1-v,v\right)$ & Saddle & $\frac{1-2v}{1-K-v}$ & $\frac{1}{3}\frac{(1-3v)}{(1-v)}$ & Depends on the point \\
		\hline
		$\mathcal{S}_{2\pm}$ & $\left(K,x,0,-x\pm\sqrt{1-K}, 1-x^2, 0\right)$ & $\begin{cases}
    & \text{Attractor for $3Q<x<0$}, \\
    & \text{Repeller for $0<x<3Q$},\\
    & \text{Saddle otherwise}. 
    \end{cases}$ & $\frac{x^2\mp 2x\sqrt{1-K}+1}{(-x\pm \sqrt{1-K})^2}$ & $\frac{1}{3}$ & Depends on the point\\
		\hline
		$\mathcal{E}_\pm$ & $\left(0,0,1,\pm\frac{1}{\sqrt{2}},0, \frac{1}{2}\right)$ & NH & $-1$ & $-1$ & De-Sitter\\
	    \hline
		$\mathcal{F}$ & $\left(\frac{3}{5},0,\frac{3}{5},0,\frac{2}{5},\frac{2}{5}\right)$ & Spiral & undefined & $-\frac{1}{3}$ & Einstein static solution\\
	    \hline
		$\mathcal{G}_\pm$ & $\left(0,\pm\frac{1}{3},\frac{8}{9},\pm\frac{2}{3},0,0\right)$ & Attractor/Repeller & $-1$ & $-1$ & De-Sitter\\
		\hline
	    $\mathcal{H}_\pm$ & $\left(\frac{2}{3},\pm\frac{1}{\sqrt{3}},\frac{2}{3},0,0,0\right)$ & Saddle & undefined& $-\frac{1}{3}$& Einstein static solution\\
	    \hline
	    $\mathcal{I}_\pm$ & $\left(\frac{3}{5},\mp\frac{1}{\sqrt{5}},\frac{4}{5},\pm\frac{1}{\sqrt{5}},0,\frac{2}{5}\right)$ & Saddle & $0$& $-\frac{1}{3}$& $a\sim t$\\
		\hline
	    \end{tabular}
		\caption{Fixed points for the HS model with the parameter choice $n=C_1=1$ in presence of CDM ($w=0$). Note that $\mathcal{S}1_\pm$ and $\mathcal{S}2_\pm$ are actually two pairs of sheets of fixed points. Deceleration parameter $q=-1-\frac{\dot{H}}{H^2}$ and the effective equation of state parameter $w_{\rm eff}=-\frac{2\dot{H}+3H^{2}+\frac{k}{a^2}}{3\left(H^{2}+\frac{k}{a^2}\right)}$ can be determined in terms of the compact dynamical variables as $q=1+\frac{K-v}{Q^2}$ and $w_{\rm eff}=\frac{Q^{2}+K-2v}{3(Q^{2}+K)}$.}
		\label{tab:fixed_ptsHS}
\end{table*}
As can be seen from the table, there are nine isolated fixed points $\mathcal{E}_\pm,\,\mathcal{F},\,\mathcal{G}_\pm,\,\mathcal{H}_\pm,\,\mathcal{I}_\pm$ and two 2-dimensional ``sheets'' of fixed points that can be expressed in a compact way as
\begin{eqnarray}
&& \mathcal{S}_1 \equiv (K,x,v,Q)  = \left((1-Q^2)\cos^{2}\theta,0,(1-Q^2)\sin^{2}\theta,Q\right),\label{S1}\\
&& \mathcal{S}_2 \equiv (K,x,v,Q) = \left(1-(Q+x)^2,x,0,Q\right).\label{S2}
\end{eqnarray}
Both $\mathcal{S}_1$ and $\mathcal{S}_2$ have one part corresponding to an expanding phase ($Q>0$) and another part corresponding to a contracting phase ($Q<0$). In Tab.\ref{tab:fixed_ptsHS} we explicitly write them as $\mathcal{S}_{1\pm}$ and $\mathcal{S}_{2\pm}$. Stability of the isolated fixed points is found in the usual way using the Hartman-Grobman theorem. Stability of the sheets of fixed points need some further discussion which we provide in Appendix \ref{app:stabilitysheets}. The entire phase space is 4-dimensional. 

The phase space of the HS model is very rich in structure and many of its features deserve an entire subsection dedicated to its detailed discussion. 
\subsection{Sheets of fixed points}
\label{subsec:sheets}
As mentioned previously, there are two pairs of sheets of fixed points $\mathcal{S}1_\pm$ and $\mathcal{S}2_\pm$ in the 4-dimensional compact phase space of the Hu-Sawicki model. Since all the points on these 2-D sheets are fixed points, there is no phase flow either on the sheet or across the sheet. Therefore each of these sheets of fixed points are also invariant submanifolds. Each of the sheets $S1_\pm$ divides the 3-dimensional submanifold $x=0$ into two disjoint volumes. Likewise, each of the sheets $S2_\pm$ divides the 3-dimensional submanifold $v=0$ into two disjoint volumes. It is to be kept in mind, however, that these sheets do \emph{not} divide the entire 4-dimensional phase space into two disjoint 4-dimensional volumes, as in that case one would require to have not a \emph{sheet} of fixed points but a \emph{volume} of fixed points (as a lower dimensional analogy one could think of a 1-parameter family of fixed points in a complete 3-dimensional phase space.).

The Jacobian of the dynamical system calculated on all four sheets yields all the four eigenvalues to be null. Therefore the Hartman-Grobman theorem cannot be used to determine their stability. We expand on this in Appendix \ref{app:stabilitysheets}.
\subsection{Physically viable regions}
\label{subsec:phys_viab}
Not the entire 4-dimensional phase space is physically viable. For the HS model with $n=C_1=1$, one can obtain from the definitions of the phase space variables $v$ and $y$ that
\begin{equation}
    C_{2}R+1=\frac{y}{v-y},
\end{equation}
and therefore 
\begin{subequations}
\begin{eqnarray}
 && f'(R)=1-\frac{1}{(C_{2}R+1)^2}=\frac{v(2y-v)}{y^2},\\
 && f''(R)=\frac{2C_2}{(C_{2}R+1)^3}=\frac{2C_{2}(v-y)^3}{y^3}.
\end{eqnarray}
\end{subequations}
Absence of ghost and tachyonic instability requires $f'>0,\,f''>0$. Keeping in mind that $C_2$ is a positive constant, this constrains the physically viable region of the phase space as
\begin{equation}
    \frac{1}{2}v < y < v,
\end{equation}
or, using the constraint equation \eqref{eq:compconstraint},
\begin{equation}
    \frac{1}{2}v < 1-K-(Q+x)^2 < v. \label{ps_const_1}
\end{equation}
Furthermore, one requires $0\leq y,\,\Omega\leq1$, which, using the constraint equations \eqref{eq:compconstraint} and \eqref{eq:friedconstraint}, translates into the conditions
\begin{eqnarray}
    && 0 \leq (Q+x)^{2}+K \leq 1, \label{ps_const_2}\\
    && 0 \leq v+x^2 \leq 1. \label{ps_const_3}
\end{eqnarray}
Constraints \eqref{ps_const_1},\eqref{ps_const_2},\eqref{ps_const_3} must be imposed additionally to single out the physically viable region of the phase space.
\subsection{Matter dominated epoch}
\label{subsec:matter}
An effectively matter dominated epoch is given by a fixed point which has $w_{\rm eff}=0$. The only candidate for this is the 1-parameter family of fixed points given by the curve
\begin{equation}
  \mathcal{M} \equiv (K,x,v,Q,\Omega,y) = \left(\frac{2}{3}-Q^{2},0,\frac{1}{3},Q,\frac{2}{3},\frac{1}{3}\right) \subset \mathcal{S}_1. \label{matt_fix}
\end{equation}
If one confines the attention to the invariant submanifold $K=0$, i.e. consider only the spatially flat cosmologies, then one gets two pairs of isolated fixed points
\begin{equation}
  \mathcal{P}_{m\pm} \equiv (x,v,Q,\Omega,y) = \left(0,\frac{1}{3},\pm\sqrt{\frac{2}{3}},\frac{2}{3},\frac{1}{3}\right).
\end{equation}
which correspond to an effectively dark matter dominated epoch during the expansion and contraction phase respectively. From Tab.\ref{tab:fixed_ptsHS} one can note that matter dominated fixed points are always saddle fixed points i.e. an intermediate epoch of a cosmology. 

\subsection{Fixed points on the $\Lambda$CDM surface}
\label{subsec:fp_LCDM}

It can be explicitly checked that the fixed points $\mathcal{E}_\pm$ and $\mathcal{G}_\pm$ satisfy the cosmographic condition \eqref{lcdm_cosm}, i.e. they fall on the $\Lambda$CDM surface. In an expanding universe, $\mathcal{E}_+$ is a saddle (i.e. an intermediate phase) whereas $\mathcal{G}_+$ is a stable fixed point (i.e. a future attractor). For the HS model under consideration to asymptotically mimic the $\Lambda$CDM evolution at late times, the phase trajectories near the matter dominated fixed point $\mathcal{M}$ must end at the De-Sitter future attractor $\mathcal{G}_+$ via the intermediate De-Sitter epoch $\mathcal{E}_+$.
\begin{figure}
    \centering
    \includegraphics[width=0.6\linewidth]{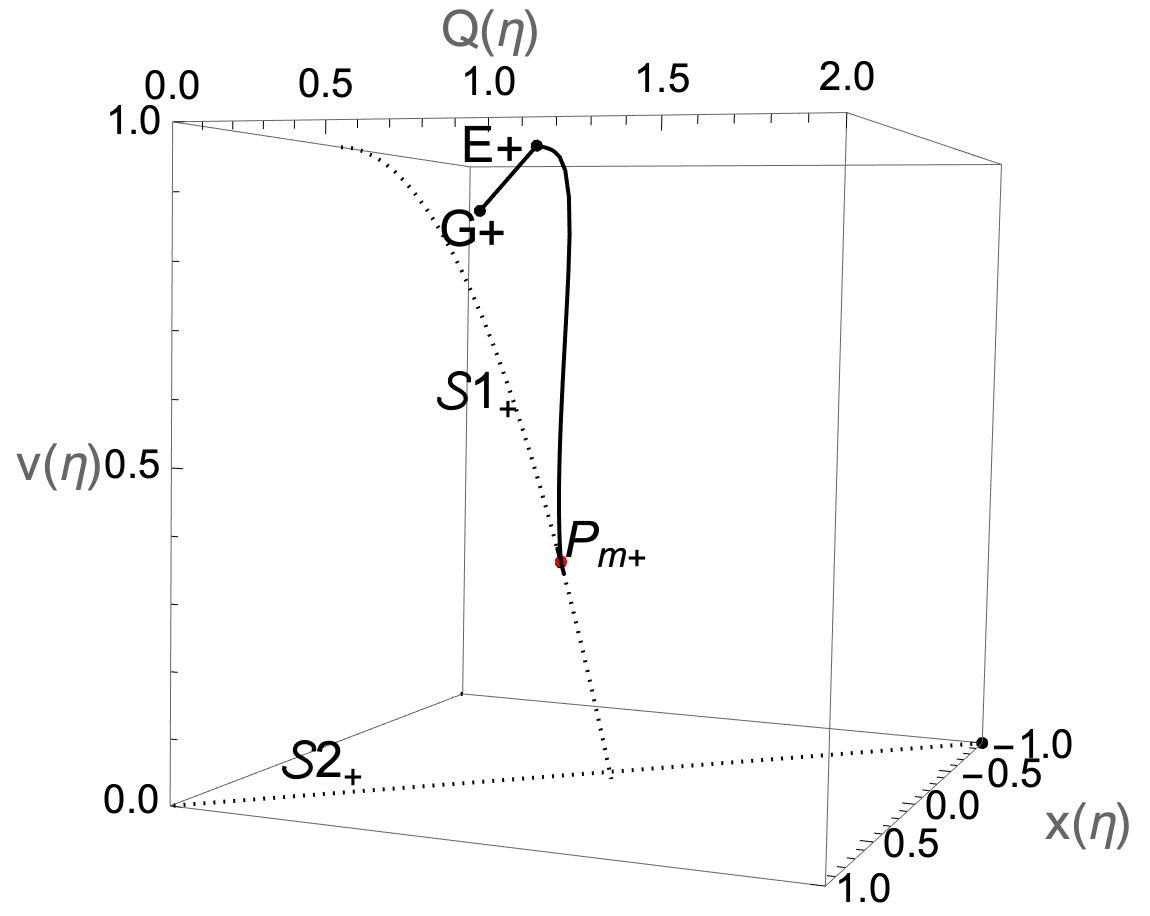}
    \caption{A characteristic phase trajectory for the spatially flat case $K=0$ passing through the point $(x,v,Q)=\left(0,0.334,\sqrt{\frac{2}{3}}\right)$ (indicated by the red dot) in the vicinity of the matter dominated fixed point $\mathcal{P}_{m+}\equiv(x,v,Q)=\left(0,\frac{1}{3},\sqrt{\frac{2}{3}}\right)$. The trajectory passes near the intermediate De-Sitter phase $\mathcal{E}_+$ and ends at the De-Sitter future attractor $\mathcal{G}_+$. Both the De-Sitter points lie on the $\Lambda$CDM surface.}
    \label{fig:currentevol_1}
\end{figure}
Fig.\ref{fig:currentevol_1} shows such a trajectory for the spatially flat case. The trajectory is plotted by numerically solving the system in eq.\eqref{eq:compactdynsysHS}, setting $\{K,w\}=\{0,0\}$ and using the initial conditions $(x,v,Q)_{\eta=0}=\left(0,0.334,\sqrt{\frac{2}{3}}\right)$, which falls in the vicinity of the matter dominated fixed point $\mathcal{P}_{m+}\equiv(x,v,Q)=\left(0,\frac{1}{3},\sqrt{\frac{2}{3}}\right)$. 

\subsection{Comparison with $\Lambda$CDM evolution}
\label{subsec:LCDM_comparison}
In this section we present the comparison of the evolution of the dimensionless Hubble parameter ($h$), deceleration parameter ($q$), effective equation of state parameter ($w_{\rm eff}$) and the jerk parameter ($j$) between the spatially flat $\Lambda$CDM cosmology and a HS cosmology given by the phase trajectory of Fig.\ref{fig:currentevol_1}, setting the same initial conditions for both. To do the numerical analysis it proves to be easier to work with the redshift $z=\frac{1-a(t)}{a(t)}$. For the spatially flat $\Lambda$CDM model the Friedmann equation and the Raychaudhuri equation are
\begin{subequations}
\begin{eqnarray}
3H^2 &=& \rho_m + \Lambda,\\
2\dot{H} + 3H^2 &=& \Lambda,
\end{eqnarray}
\end{subequations}
from which one can write the evolution of the dimensionless Hubble parameter and the deceleration parameter as\footnote{For the spatially flat $\Lambda$CDM model, the jerk parameter $j$ is identically unity \cite{Dunajski:2008tg}}
\begin{subequations}\label{lcdm_evoln}
\begin{eqnarray}
&& h(z) = \sqrt{\Omega_{m0}(1+z)^3 + (1-\Omega_{m0})},\\
&& q(z) = \frac{3\Omega_{m0}(1+z)^3}{2h^{2}(z)} - 1,\\
&& w_{\rm eff}(z) = - \frac{1}{3}(1 - 2q(z)),\\
&& j(z) = 1.
\end{eqnarray}
\end{subequations}

Noting that $dN=-\frac{dz}{1+z}$, one can express the non-compact dynamical system in Eq.\eqref{dynsys_1} with respect to $z$:
\begin{subequations}\label{NCdynsys_1}
\begin{align}
\frac{d\tilde{x}}{dz}  =& -\frac{1}{(z+1)}\Big[-2\tilde{v}+4+3\tilde{x}-3(1+w)\tilde{\Omega}+4\tilde{K}-\tilde{x}^2-\tilde{x}\tilde{v}+\tilde{x}\tilde{K}\Big],\\
      \frac{d\tilde{v}}{dz} =& -\frac{\tilde{v}}{(z+1)}\left(\Gamma \tilde{x}-2\tilde{v}+2\tilde{K}+4\right),\\
\frac{d\tilde{K}}{dz}=&\frac{2\tilde{K}}{(z+1)}\left(\tilde{v}-\tilde{K}-1\right),\\
  \frac{d\tilde{\Omega}}{dz}
    =& \frac{\tilde{\Omega}}{(z+1)}\left(-1+3w+\tilde{x}+2\tilde{v}-2\tilde{K}\right),\\
     \frac{d\tilde{y}}{dz}
    =& - \frac{1}{(z+1)}\left(\Gamma \tilde{x}\tilde{v} + \tilde{y}(2\tilde{K}-2\tilde{v}-\tilde{x}+4)\right).
\end{align}
\end{subequations}
For the spatially flat case $\tilde{K}=0$, the constraint equation \eqref{constr_1} gives
\begin{equation}
    \tilde{\Omega} = 1 + \tilde{x} + \tilde{y} - \tilde{v},
\end{equation}
and the non-compact dynamical system \eqref{NCdynsys_1} reduces to
\begin{subequations}\label{NCdynsys_2}
\begin{align}
\frac{d\tilde{x}}{dz}  =& -\frac{1}{(z+1)}\Big[-2\tilde{v}+4+3\tilde{x}-3(1+w)(1+\tilde{x}+\tilde{y}-\tilde{v}) -\tilde{x}^2-\tilde{x}\tilde{v}\Big],
\\
\frac{d\tilde{v}}{dz} =& -\frac{\tilde{v}}{(z+1)}\left(\Gamma \tilde{x}-2\tilde{v}+4\right),
\\
 \frac{d\tilde{y}}{dz}
    =& - \frac{1}{(z+1)}\left(\Gamma \tilde{x}\tilde{v} + \tilde{y}(-2\tilde{v}-\tilde{x}+4)\right).
\end{align}
\end{subequations}

To solve the system in Eq.\eqref{NCdynsys_2}, one needs to set an initial condition $(\tilde{x},\tilde{v},\tilde{y})_{z_i}$ at some $z_i$. To compare the evolution of the spatially flat $\Lambda$CDM model and the spatially flat HS model with $n=1=C_1$, one needs to set the same initial condition for the two models at some time in the past. Following the idea adopted in Ref.\cite{Kandhai:2015pyr}, we set the initial condition at $z=20$. In the remainder of this section we explain how to set the same initial condition for both the cosmologies at $z=20$.

Taking the parameter choice $(\Omega_{m0},\Omega_{\Lambda 0})=(0.315,0.685)$\footnote{Actual Planck values are $(\Omega_{\Lambda0},\Omega_{k0},\Omega_{m0})=(0.6843\pm0.0073,0.0007\pm0.0019,0.315\pm0.007)$ \cite{Planck2020}. However, since we are comparing between spatially flat cosmologies, it is a good enough approximation to take $(\Omega_{m0},\Omega_{\Lambda 0})=(0.315,0.685)$.} in Eq.\eqref{lcdm_evoln}, for the $\Lambda$CDM evolution one gets
\begin{equation}
(h,q)_{z=20} = (h_{\Lambda \rm CDM},q_{\Lambda \rm CDM})_{z=20} = (54.0176,0.499).  
\end{equation}
Setting $w=0$ (CDM) and $\Gamma=\frac{1}{2}\frac{\tilde{v}\tilde{y}}{(\tilde{v}-\tilde{y})^2}$ (for HS model with $n=1=C_1$) and using the initial conditions, one can numerically solve the system \eqref{NCdynsys_1} to find the evolution of various dynamical quantities along the trajectory. Evolution of different cosmographic parameters and the effective equation of state parameters are then obtained as
\begin{subequations}
\begin{eqnarray}
q(z) &=& - 1 - \frac{\dot{H}}{H^2} = 1 - \tilde{v}(z),\label{eq:decelparam}\\
w_{\textnormal{eff}}(z) &=& - \frac{2\dot{H} + 3H^2}{3H^2} = \frac{1}{3}(1-2\tilde{v}(z)),\\
j(z) &=& 2q^2(z) + q(z) - \frac{dq(z)}{dN} = 2q^2(z) + q(z) + (1+z)\frac{dq(z)}{dz},\\
h(z) &=& \exp\left[\int_0^z \frac{1+q(z)}{1+z}dz\right].
\end{eqnarray}
\end{subequations}
One now gets from \eqref{eq:decelparam} that
\begin{equation}\label{IC_vtilde}
    \tilde{v}(z=20) = 1 - q(z=20) = 0.501.
\end{equation}
Since we are setting the same initial conditions for the $\Lambda$CDM and HS cosmology at $z=20$, we assume that gravity at $z=20$ is described well enough by GR, so that
\begin{equation}\label{IC_xtilde}
    \tilde{x}(z=20) = 0.
\end{equation}

One still needs to determine the initial condition $\tilde{y}(z=20)$. To this goal, we adopt the following procedure. The past asymptotic of the $\Lambda$CDM model is the matter dominated phase; $\Omega_m \gg \Omega_\Lambda$ for $z\gg1$. We consider the $f(R)$ cosmology to be at the matter dominated phase at $z=20$, so that
\begin{equation}
    Q(z=20) = Q\vert_{\mathcal{P}_{m+}} = \sqrt{\frac{2}{3}},
    \label{eq:QPm}
\end{equation}
$\mathcal{P}_{m+}$ being the matter dominated fixed point. Setting $K=0$ (since we are considering spatially flat cosmologies here) and $x(z=20)=0$ (since $\tilde{x}(z=20)=0$), $y(z=20)$ can be found from the constraint equation \eqref{eq:compconstraint} to be
\begin{equation}
    y(z=20) = \frac{1}{3}.
\end{equation}
One now gets from Eq.\eqref{var_reln} that
\begin{equation}\label{IC_ytilde}
    \tilde{y}(z=20) = \frac{y(z=20)}{Q^{2}(z=20)} = \frac{1}{2}.
\end{equation}
Eqs.\eqref{IC_vtilde},\eqref{IC_xtilde},\eqref{IC_ytilde} provides the initial conditions necessary to solve the system in eq.\eqref{NCdynsys_1}.

Once $Q(z=20)$ is known, using Eq.\eqref{var_reln} one can obtain the numerical values of the compact dynamical variables at $z=20$ from the numerical values of the corresponding noncompact dynamical variables at $z=20$:
\begin{equation}
    \left(x,v,Q\right)_{z=20} = \left(Q^2 \tilde{x}, Q^2 \tilde{v}, Q\right)_{z=20} \left(0,0.334,\sqrt{\frac{2}{3}}\right),
    \label{eq:initconditLCDM}
\end{equation}
In the compact phase space this point lies in the vicinity of the matter dominated fixed point $\mathcal{P}_{m+}$. The phase trajectory shown in Fig.\ref{fig:currentevol_1} passes through this point. Solving the system in Eq.\eqref{NCdynsys_1} with the initial conditions given by Eqs.\eqref{IC_vtilde},\eqref{IC_xtilde},\eqref{IC_ytilde} we find the evolution of various quantities along this particular trajectory.

\begin{figure*}
    \centering
    \subfigure[]{\includegraphics[width=0.3\linewidth]{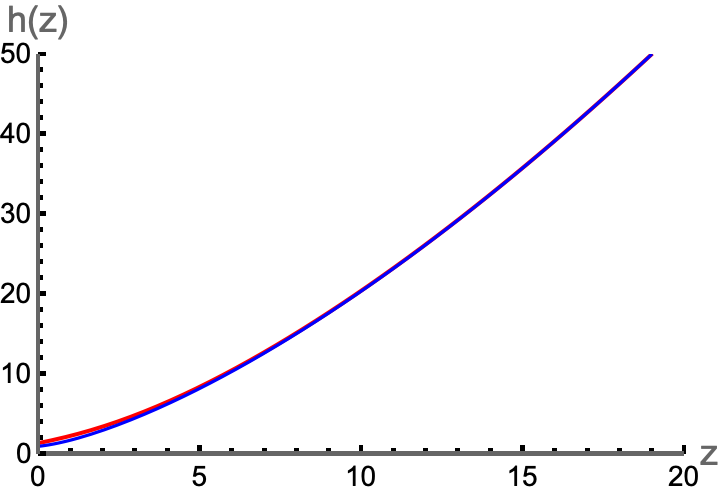}}
    \hspace{0.3cm}
    \subfigure[]{\includegraphics[width=0.3\linewidth]{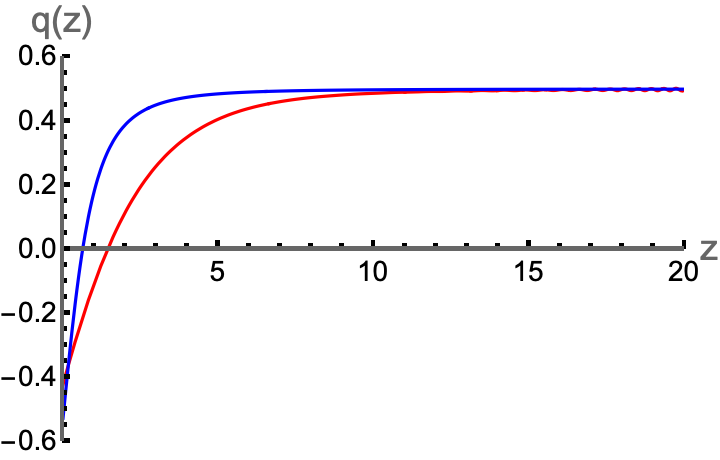}}
    \hspace{0.3cm}
    \subfigure[]{\includegraphics[width=0.3\linewidth]{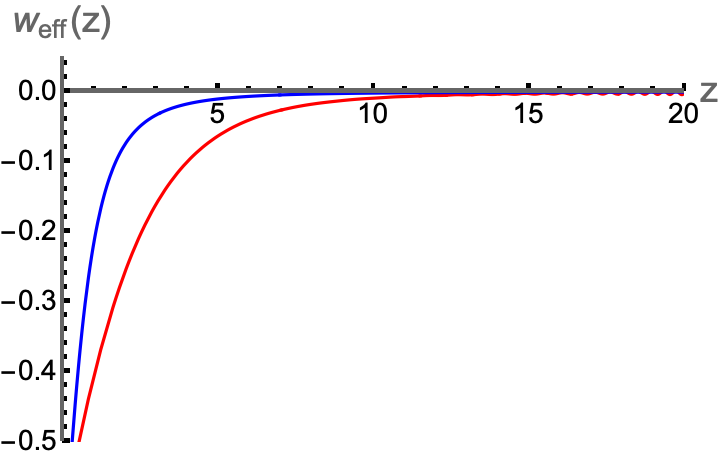}}
    \caption{Evolution with respect to the redshift ($z=\frac{1-a}{a}$) of the dimensionless Hubble parameter ($h$), deceleration parameter ($q$) and effective equation of state parameter ($w_{\rm eff}$) along the phase trajectory of Fig.\ref{fig:currentevol_1} (red curve) and the corresponding evolutions for the $\Lambda$CDM model (blue curve). The initial conditions for the non-compact variables, corresponding to those used in Fig.\ref{fig:currentevol_1}, are $(\tilde{x},\tilde{v},\tilde{y})=(0,0.501,0.5)$ and are set at $z=20$ following \cite{Kandhai:2015pyr}.}
    \label{fig:currentevol_2}
\end{figure*}
\begin{figure*}
    \centering
   \includegraphics[width=0.8\textwidth]{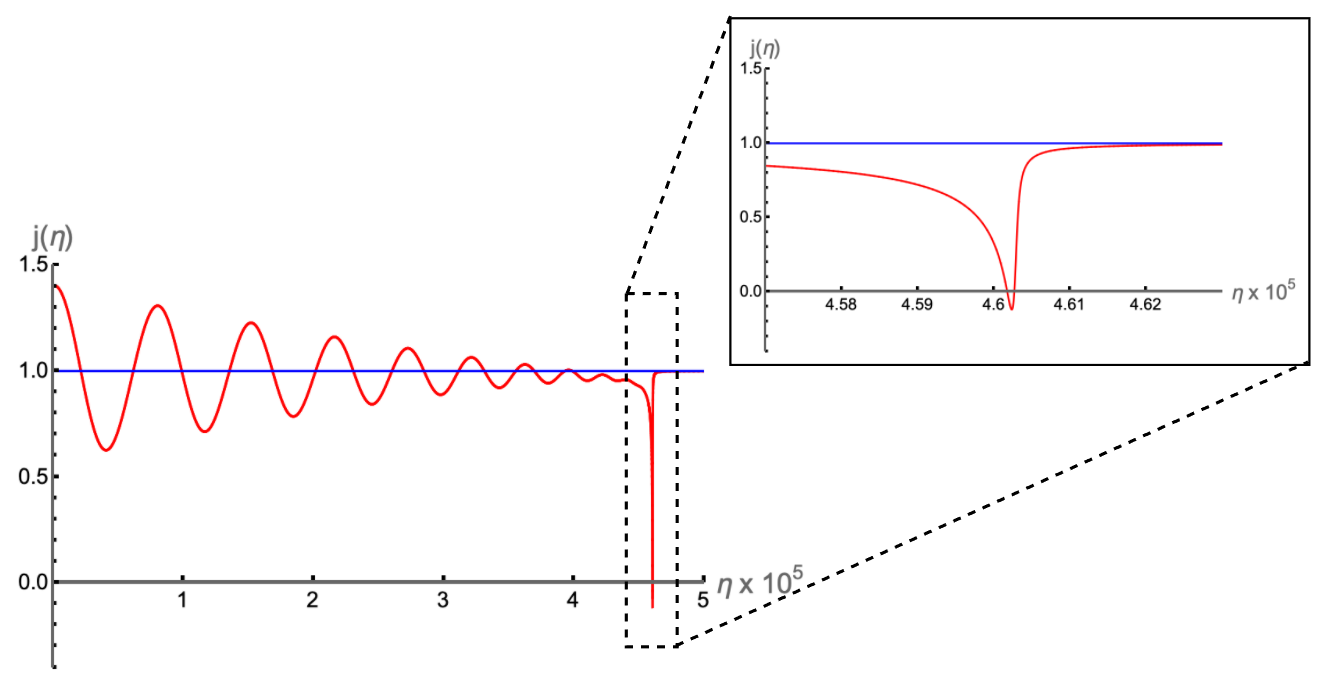}
    \caption{Evolution with respect to the redshift ($z=\frac{1-a}{a}$) of the jerk parameter ($j$) along the phase trajectory of Fig.\ref{fig:currentevol_1} (red curve) and the corresponding evolutions for the $\Lambda$CDM model (blue curve). The initial conditions for the non-compact variables, corresponding to those used in Fig.\ref{fig:currentevol_1}, are $(\tilde{x},\tilde{v},\tilde{y})=(0,0.501,0.5)$ and are set at $z=20$ following \cite{Kandhai:2015pyr}. The plot reaffirms the fact that the phase trajectory of Fig.\ref{fig:currentevol_1} passes through the De-Sitter saddle $\mathcal{E}_+$ and settles ultimately on the De-Sitter attractor $\mathcal{G}_+$, both of which fall on the surface $j=1$ (see subsection \ref{subsec:fp_LCDM}).}
    \label{fig:jLCDM}
\end{figure*}

In Fig.\ref{fig:currentevol_2}, we show the evolution of the dimensionless Hubble parameter, deceleration parameter and the effective equation of state for the trajectory of Fig.\ref{fig:currentevol_1} with respect to the redshift $z$, and compare it to the $\Lambda$CDM evolution. 
In Fig.\ref{fig:jLCDM} we show the evolution of the jerk parameter with respect to the phase space time variable $\eta$ defined in Eq.\eqref{time_redef} and compare with that of the $\Lambda$CDM model (i.e. $j=1$).

\subsection{Nonsingular bounce, recollapse and cyclic cosmology}
\label{sec:bounces}
\begin{figure*}
    \centering
    \subfigure[\label{fig:kneq02Da}]{\includegraphics[width=0.3\linewidth]{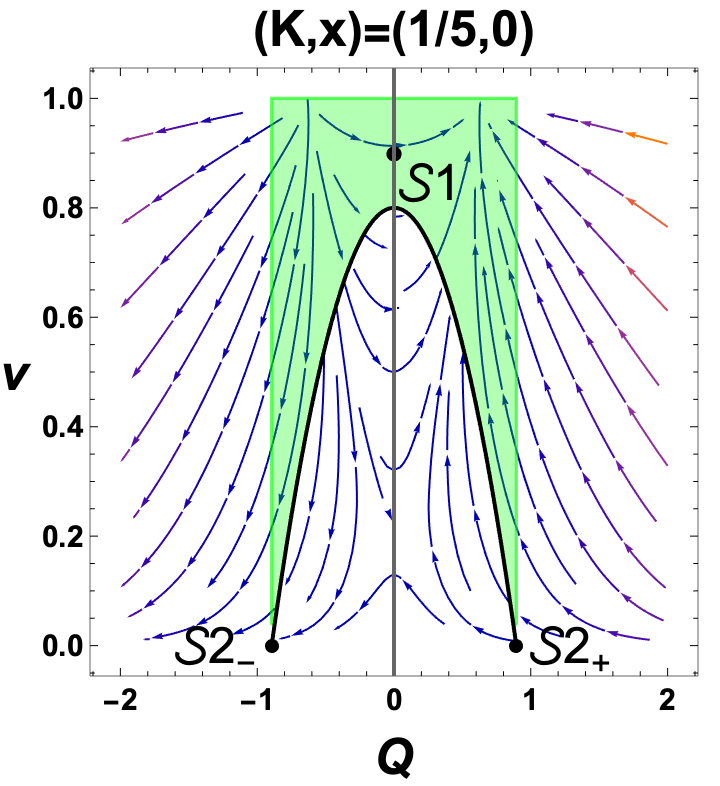}}
    \hspace{0.3cm}
    \subfigure[\label{fig:kneq02Db}]{\includegraphics[width=0.3\linewidth]{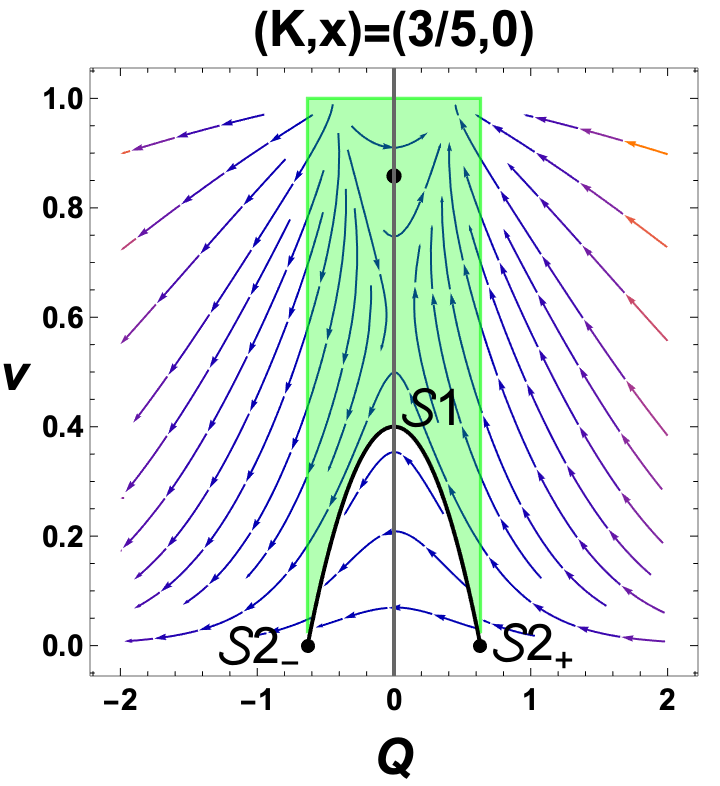}}
    \hspace{0.3cm}
    \subfigure[\label{fig:kneq02Dc}]{\includegraphics[width=0.3\linewidth]{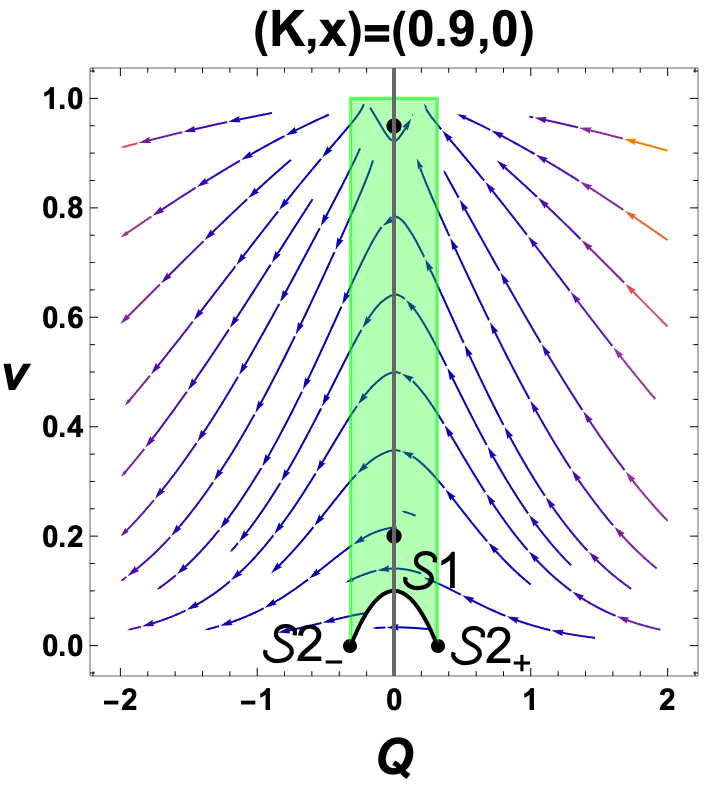}}
    \caption{2-D slices of the phase space for various values of $K\neq 0$ and $x=0$. The sheet of fixed points $\mathcal{S}_1$ intersects these 2-D slices at curves that are represented by the black curves in the plots. The sheet of fixed points $\mathcal{S}_2$ intersects these 2-D slices only at the two points $(Q,v)=(\pm\sqrt{1-K},0)$, which, coincidentally, also fall on $\mathcal{S}_1$. Increasing the value of $K$ allows more trajectories to pass through $Q=0$ as either bounces or recollapses. The green shaded region corresponds to the subset of the three phase space viability constraints \eqref{ps_const_1},\eqref{ps_const_2},\eqref{ps_const_3}, as projected onto the 2-D $Q$-$v$ slices. }
    \label{fig:kneq02D}
\end{figure*}

A nonsingular bounce and a recollapse can be specified respectively by the conditions
\begin{equation}
    \dot{H}\vert_{H=0}>0, \qquad \dot{H}\vert_{H=0}<0.
\end{equation}
Since $Q=\frac{3H}{D}=0$ is zero for a nonsingular bounce or a recollapse, a phase trajectory corresponding to a bounce or a recollapse must cross the $Q=0$ hypersurface. For a nonsingular bounce ($\dot{H}>0$) the crossing is from $Q<0$ side to $Q>0$ side and for a recollapse ($\dot{H}<0$) the crossing is from $Q>0$ side to $Q<0$ side.
Starting from the definition of the Ricci scalar, one can express $\dot{H}$ as 
\begin{eqnarray}
\dot{H}&&=\frac{D^2}{9}\left(v-2Q^2-K\right).
\end{eqnarray}
 Using the above expression, one can distinguish the regions on the $Q=0$ hypersurface in which the trajectories cross from left to right (bounce) and from right to left (recollapse) as well as the region in which the trajectories don't cross at all. They are respectively
 \begin{equation}\label{bounce_cond}
     \begin{aligned}
     & v>K \qquad\qquad (\text{Bounce})\\
     & v<K \qquad\qquad (\text{Recollapse})\\
     & v=K \qquad\qquad (\text{No crossing}).
     \end{aligned}
 \end{equation}
On the $Q=0$ hypersurface the physical viability conditions \eqref{ps_const_1},\eqref{ps_const_2},\eqref{ps_const_3} become
\begin{equation}
    \frac{1}{2}v < 1-K-x^2 < v, \qquad 0 \leq K+x^2 \leq 1, \qquad 0\leq v+x^2 \leq 1.
\end{equation}
Clearly, for the spatially flat case $K=0$, the first and the third inequality are incompatible. This implies a breakdown of the condition $f''>0$. One therefore arrives at an important conclusion:
\begin{itemize}
    \item No physically viable nonsingular bounce or recollapse is possible in spatially flat FLRW cosmology for the Hu-Sawicki model with $n=1=C_1$.
\end{itemize}
Note that this conclusion holds only for $K=0$. For $K\neq0$ perfectly viable bounces or recollapses can occur. Increasing the value of $K$ from $0$ to $1$ allows for more space for the first and third inequality to be simultaneously satisfied, and consequently increases the possibility of a perfectly viable bounce (see 2-dimensional slices of the phase space in Fig.\ref{fig:kneq02D}). 

\begin{figure*}
    \centering
    \subfigure[\label{fig:cyclica}]{\includegraphics[width=0.3\linewidth]{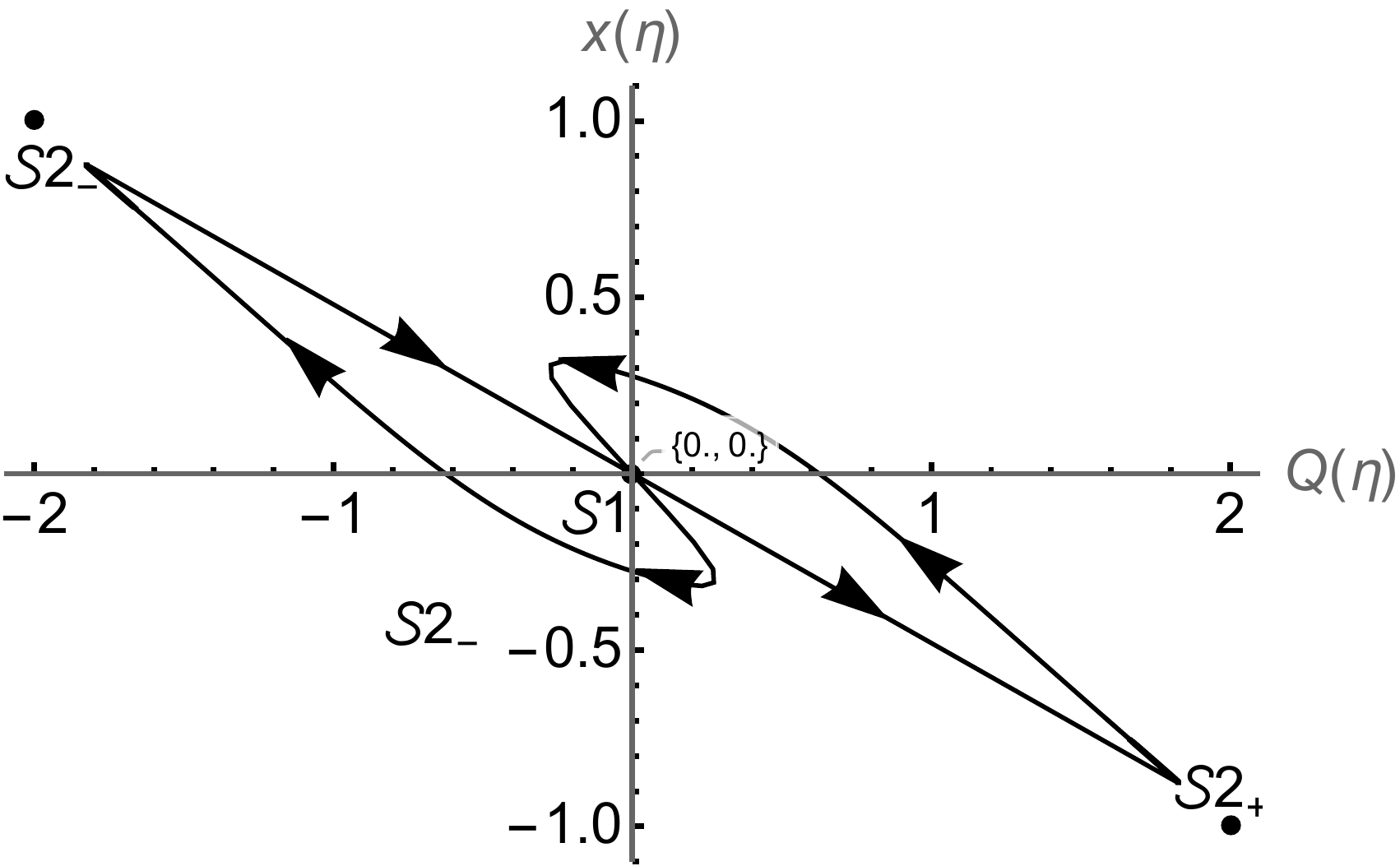}}
    \hspace{0.3cm}
    \subfigure[\label{fig:cyclicb}]{\includegraphics[width=0.3\linewidth]{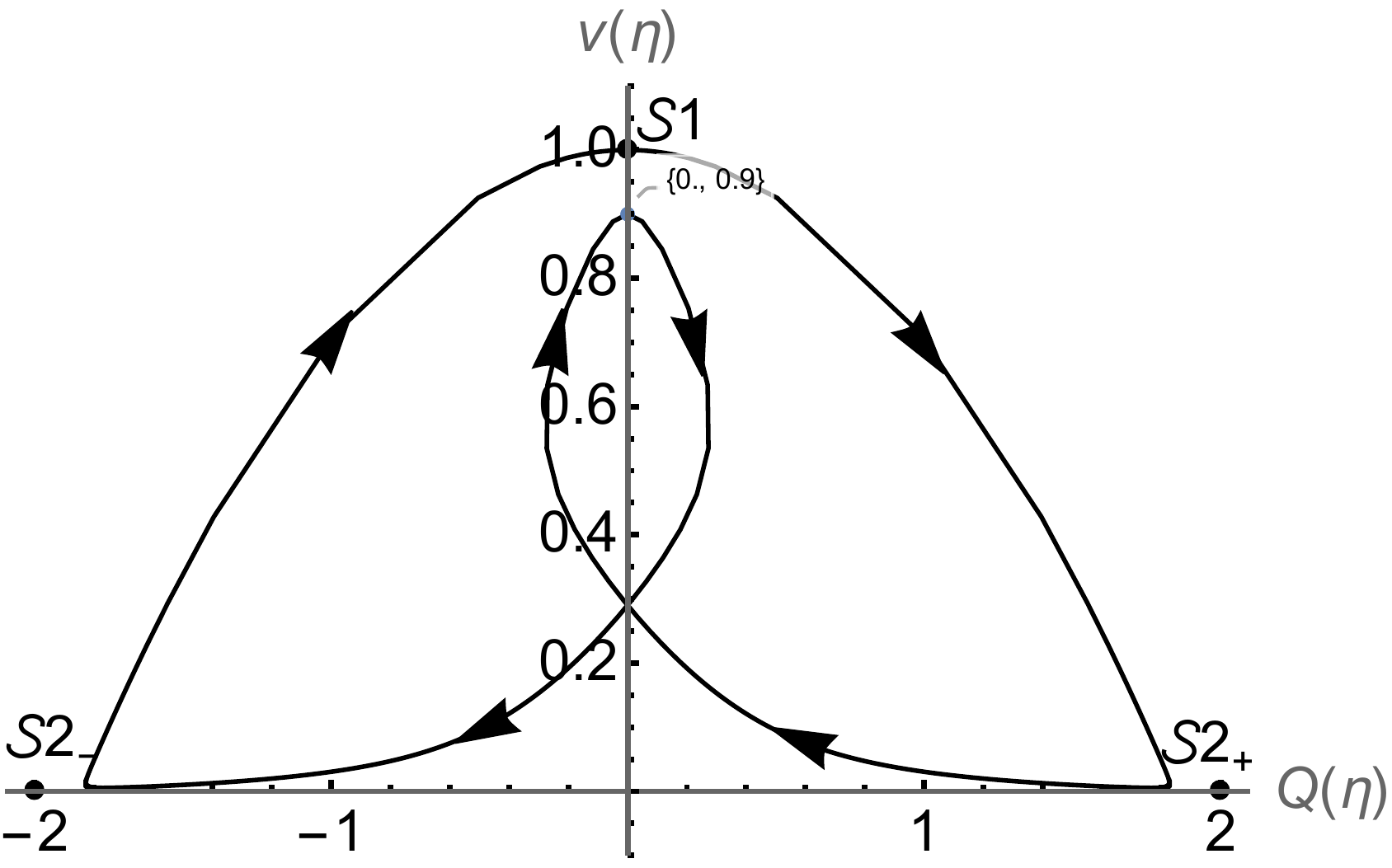}}
    \hspace{0.3cm}
    \subfigure[\label{fig:cyclicc}]{ \includegraphics[width=0.3\linewidth]{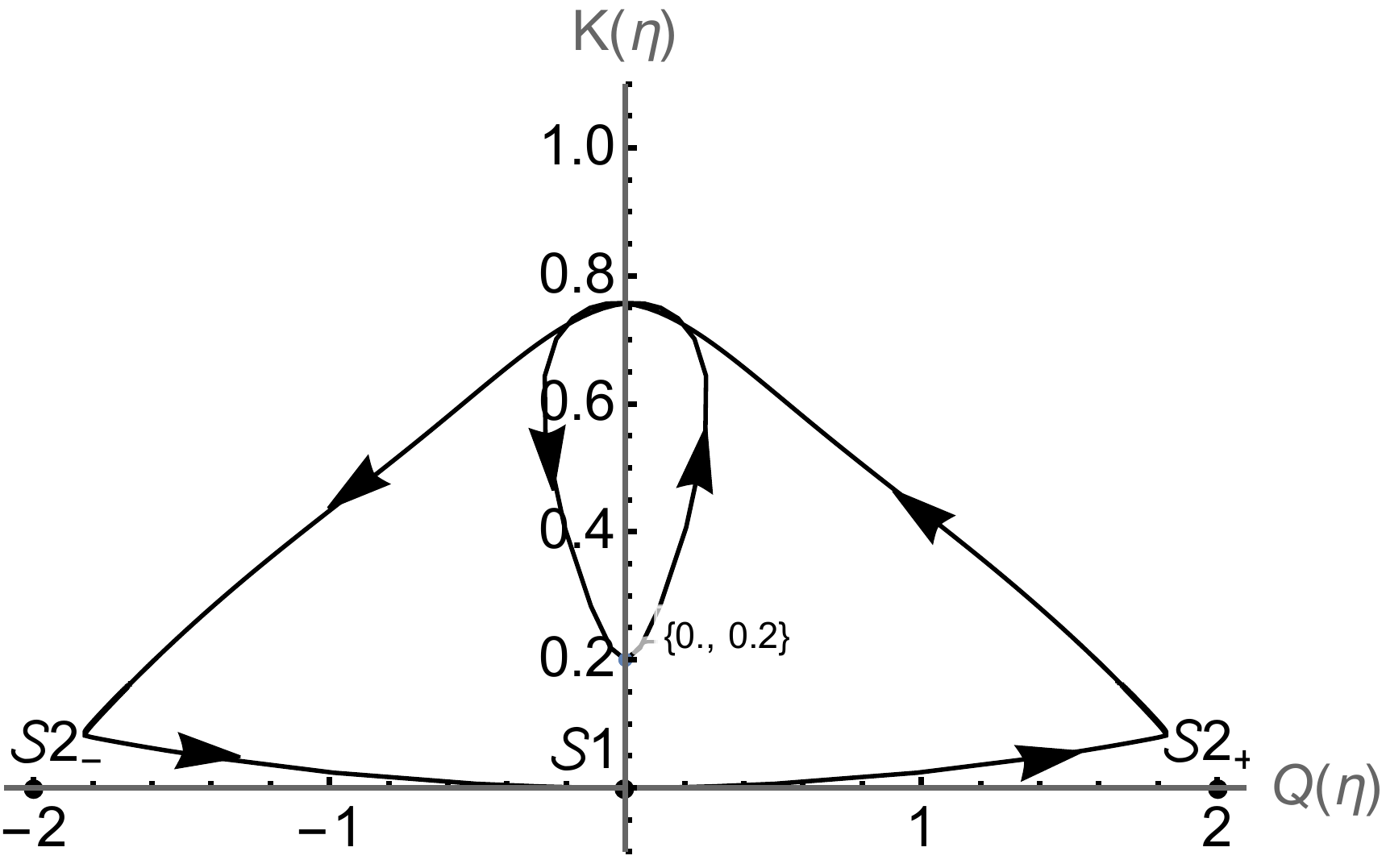}}
    \caption{Parametric plots corresponding to the initial conditions $(K(0),x(0),v(0),Q(0))=(0.2,0,0.9,0)$ (coordinates labelled). The corresponding point in the phase space is marked with a blue dot in Fig.\ref{fig:kneq02Da}. The trajectory evolves away from the saddle fixed point $(K,x,v,Q)=(0,0,1,0)$ on $\mathcal{S}_{1}$ in the past towards the saddle fixed point $(K,x,v,Q)=(0,-1,0,2)$ on $\mathcal{S}_{2+}$. It then undergoes a recollapse in the past and a bounce at the initial condition point $(K,x,v,Q)=(0.2,0,0.9,0)$ (since $v>K$ at this point; see Eq.\eqref{bounce_cond}) then recollapses again and evolves towards the corresponding collapsing saddle fixed point $(K,x,v,Q)=(0,1,0,-2)$ on $\mathcal{S}_{2-}$ in the future, before returning to the point of origin on $\mathcal{S}_{1}$. This entire cycle is repeated over and over again and this cyclic trajectory corresponds to a cyclic cosmology. We note here that the recollapse points look to be identical from Fig.\ref{fig:cyclicb} and \ref{fig:cyclicc}, however the $x$ coordinate switches sign through the bounce between them, as seen in Fig.\ref{fig:cyclica}.}
    \label{fig:k1/5parametric}
\end{figure*}

\begin{figure*}
    \centering
    \subfigure[]{\includegraphics[width=0.3\linewidth]{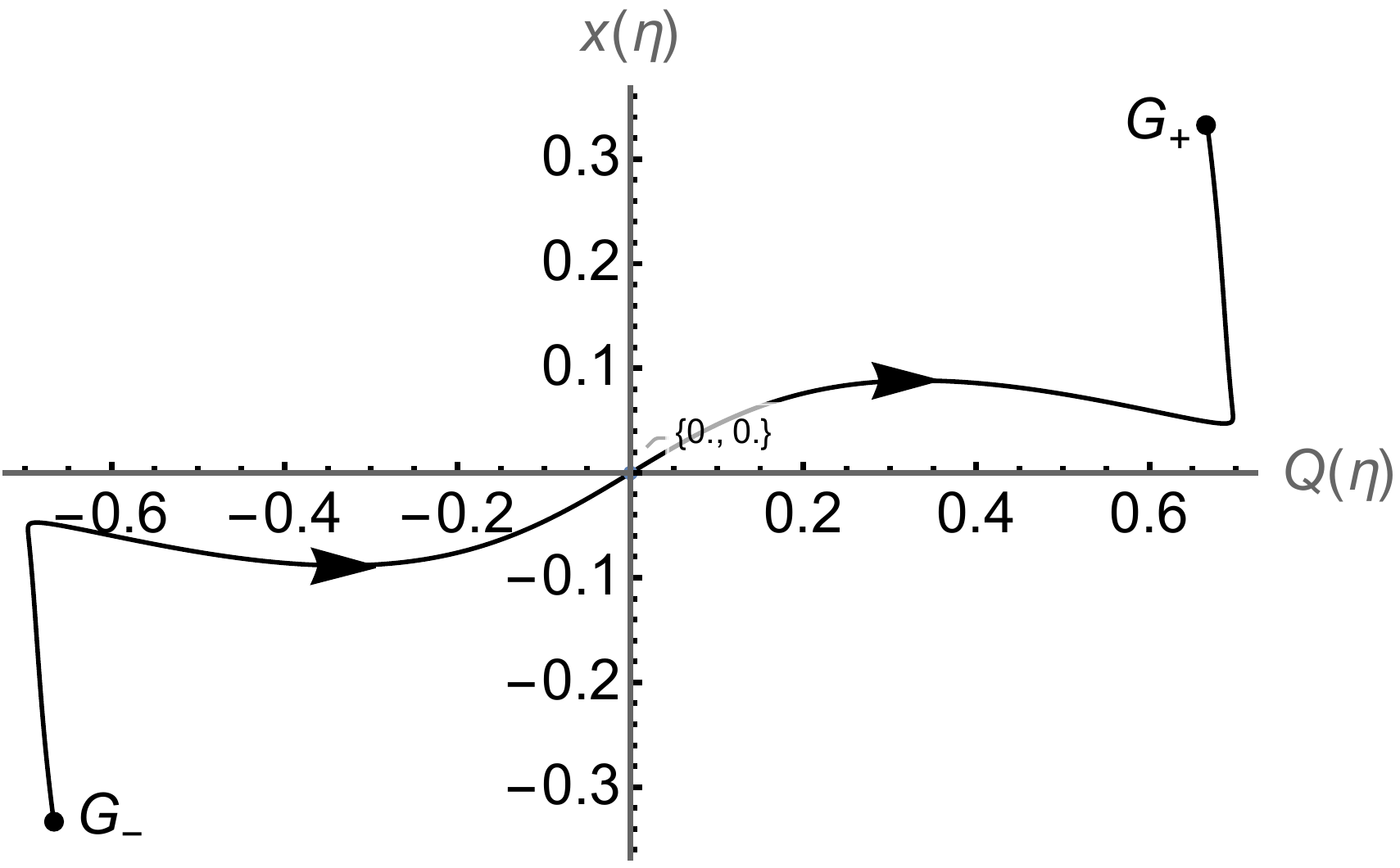}}
    \hspace{0.3cm}
    \subfigure[]{\includegraphics[width=0.3\linewidth]{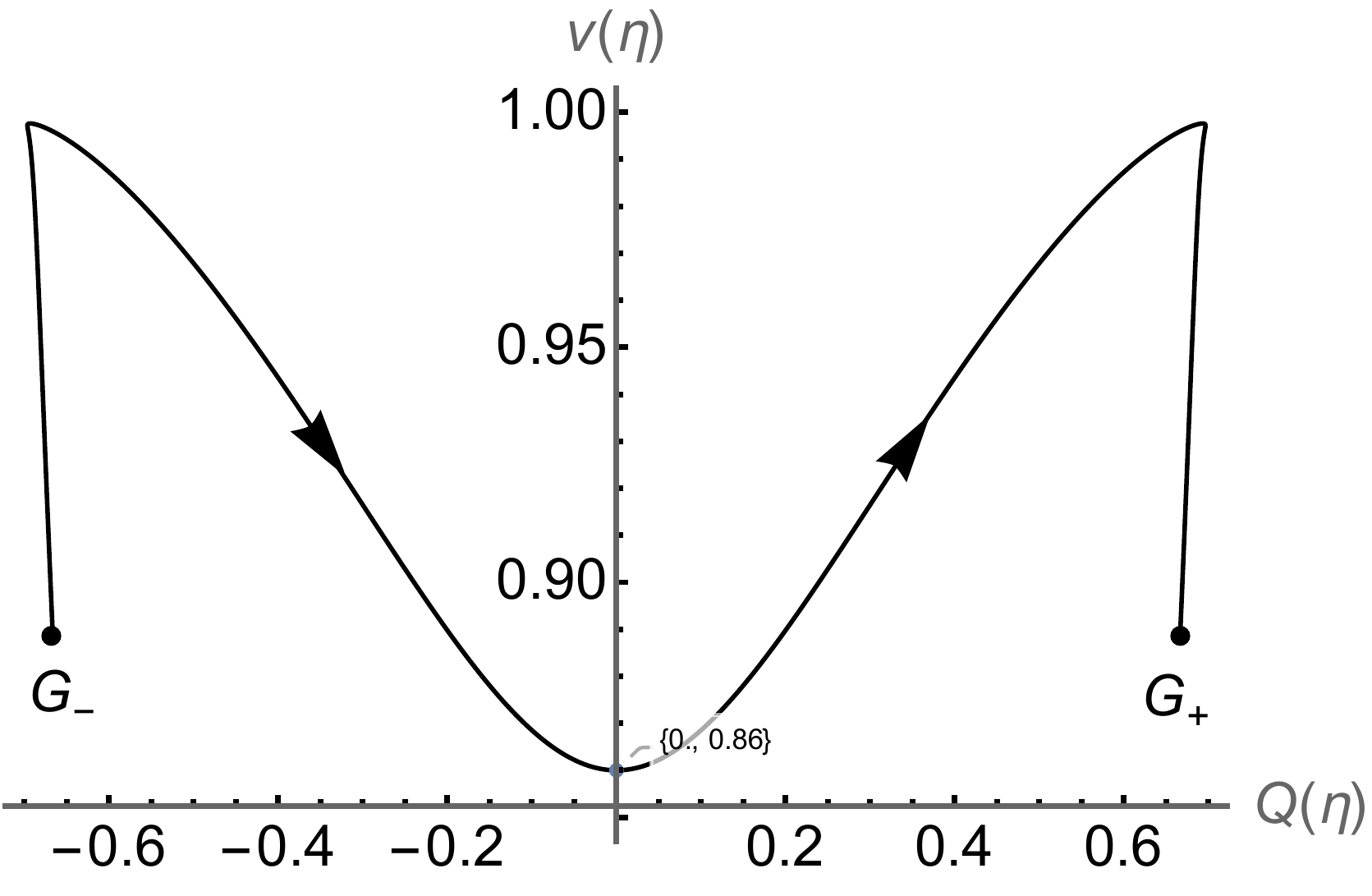}}
    \hspace{0.3cm}
    \subfigure[]{\includegraphics[width=0.3\linewidth]{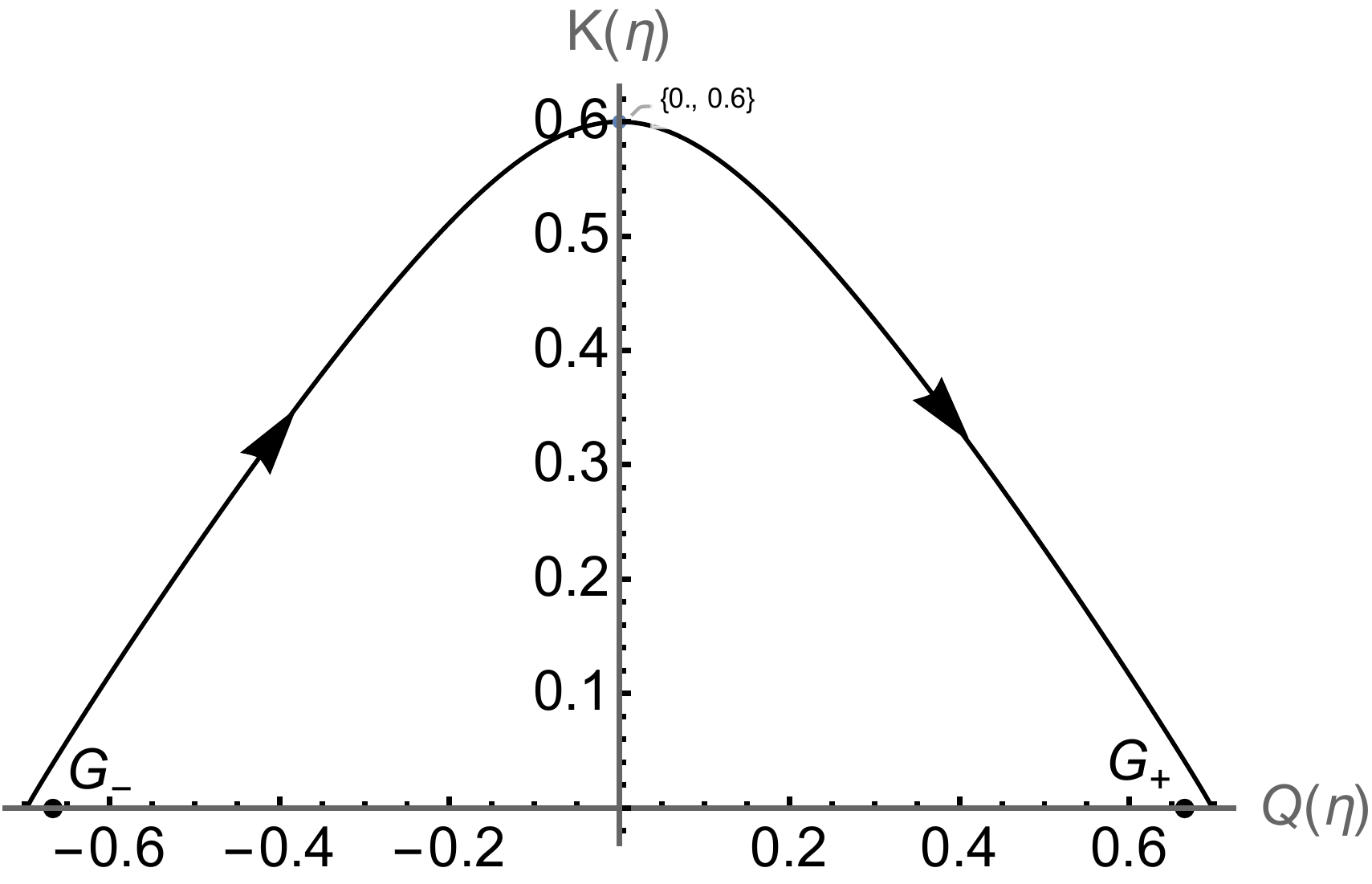}}
    \caption{Parametric plots corresponding to the initial conditions $(K(0),x(0),v(0),Q(0))=(0.6,0,0.86,0)$ (coordinates labelled). The corresponding point in the phase space is marked with a blue dot in Fig.\ref{fig:kneq02Db}. The trajectory evolves asymptotically from the past attractor $\mathcal{G}_-$ to the future attractor $\mathcal{G}_+$, undergoing a bounce at $(K,x,v,Q)=(0.6,0,0.86,0)$ (since $v>K$ at this point; see Eq.\eqref{bounce_cond}).}
    \label{fig:k3/5parametric}
\end{figure*}

\begin{figure*}
    \centering
    \subfigure[]{\includegraphics[width=0.3\linewidth]{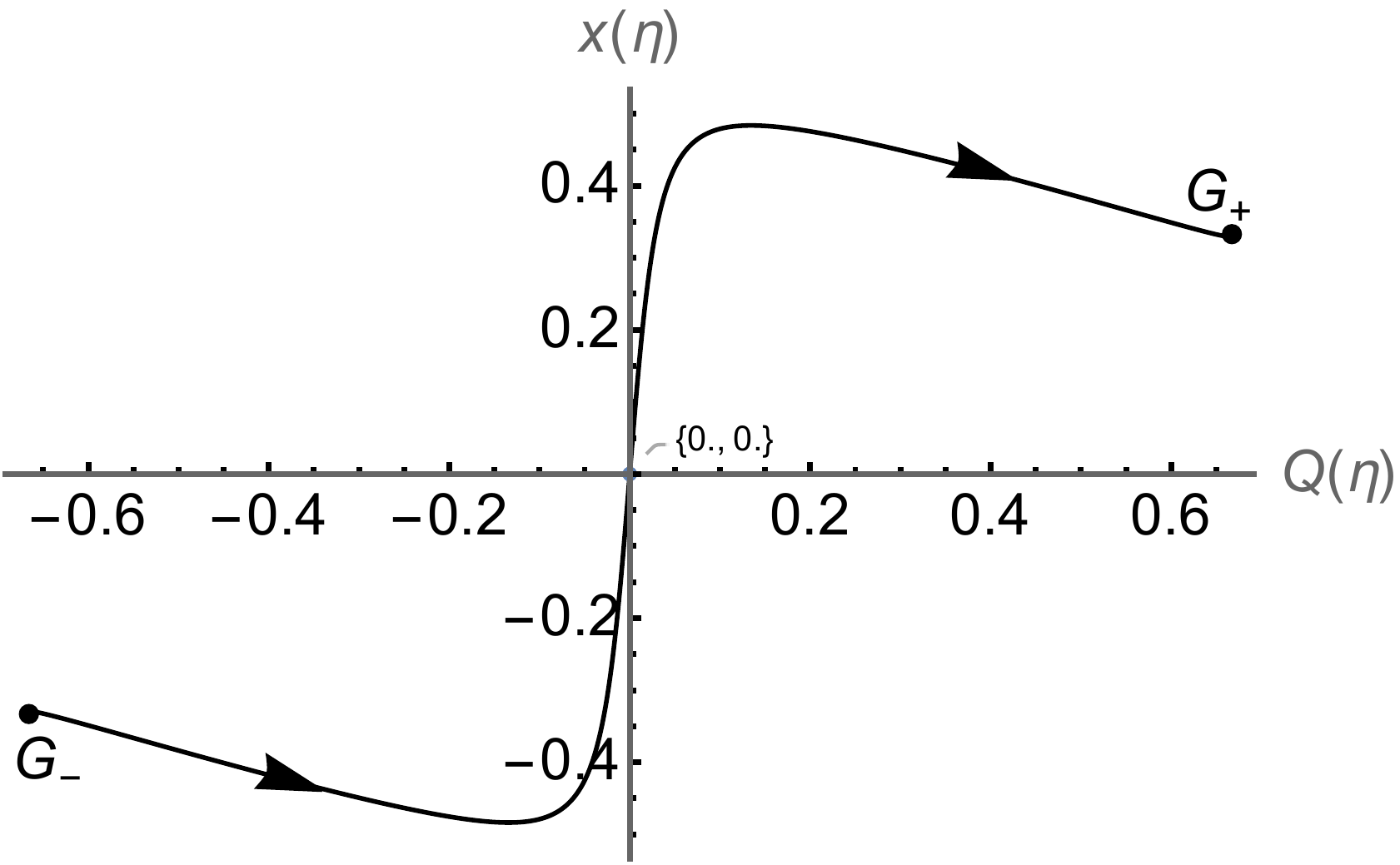}}
    \hspace{0.3cm}
    \subfigure[]{\includegraphics[width=0.3\linewidth]{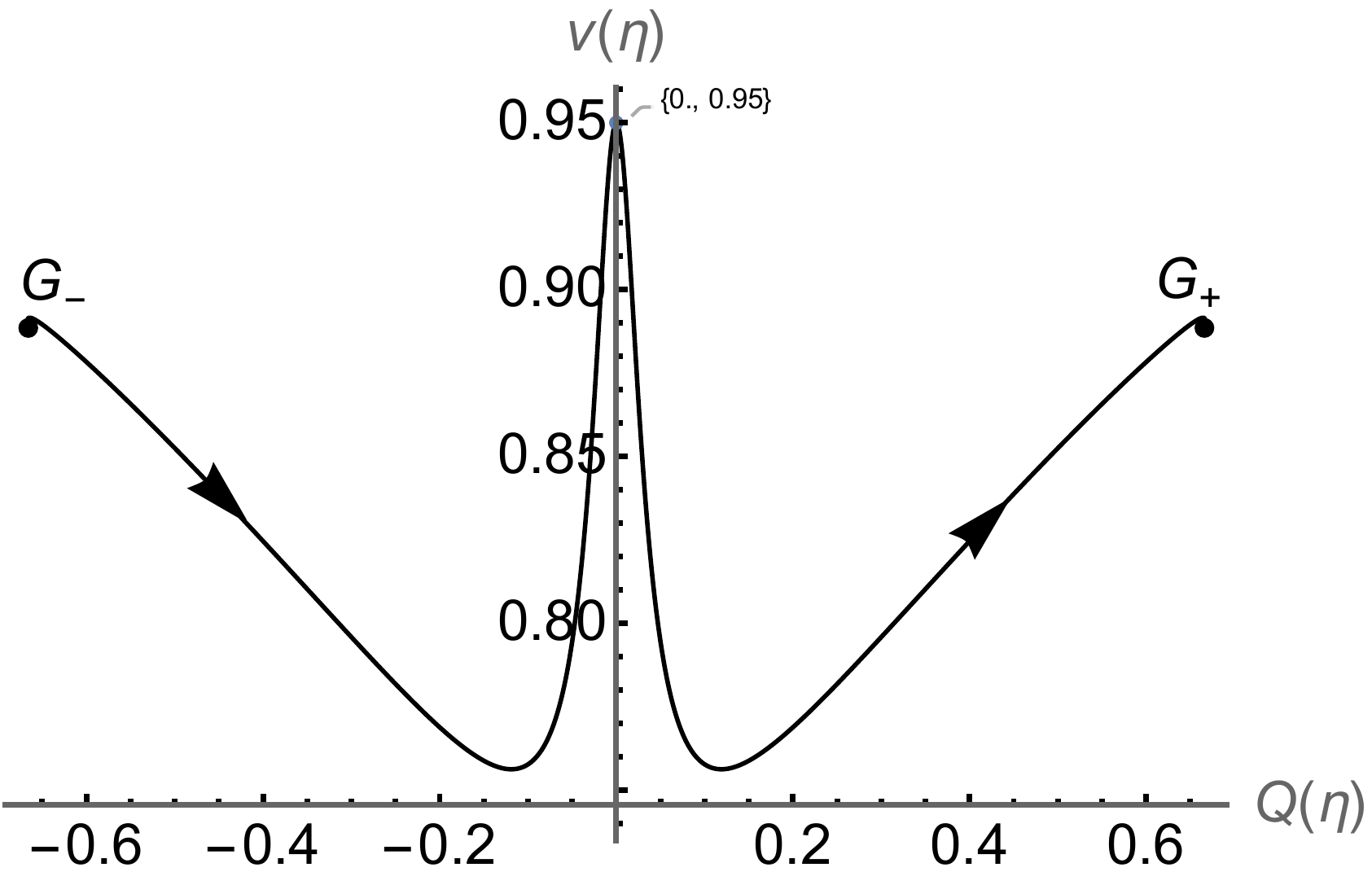}}
    \hspace{0.3cm}
    \subfigure[]{\includegraphics[width=0.3\linewidth]{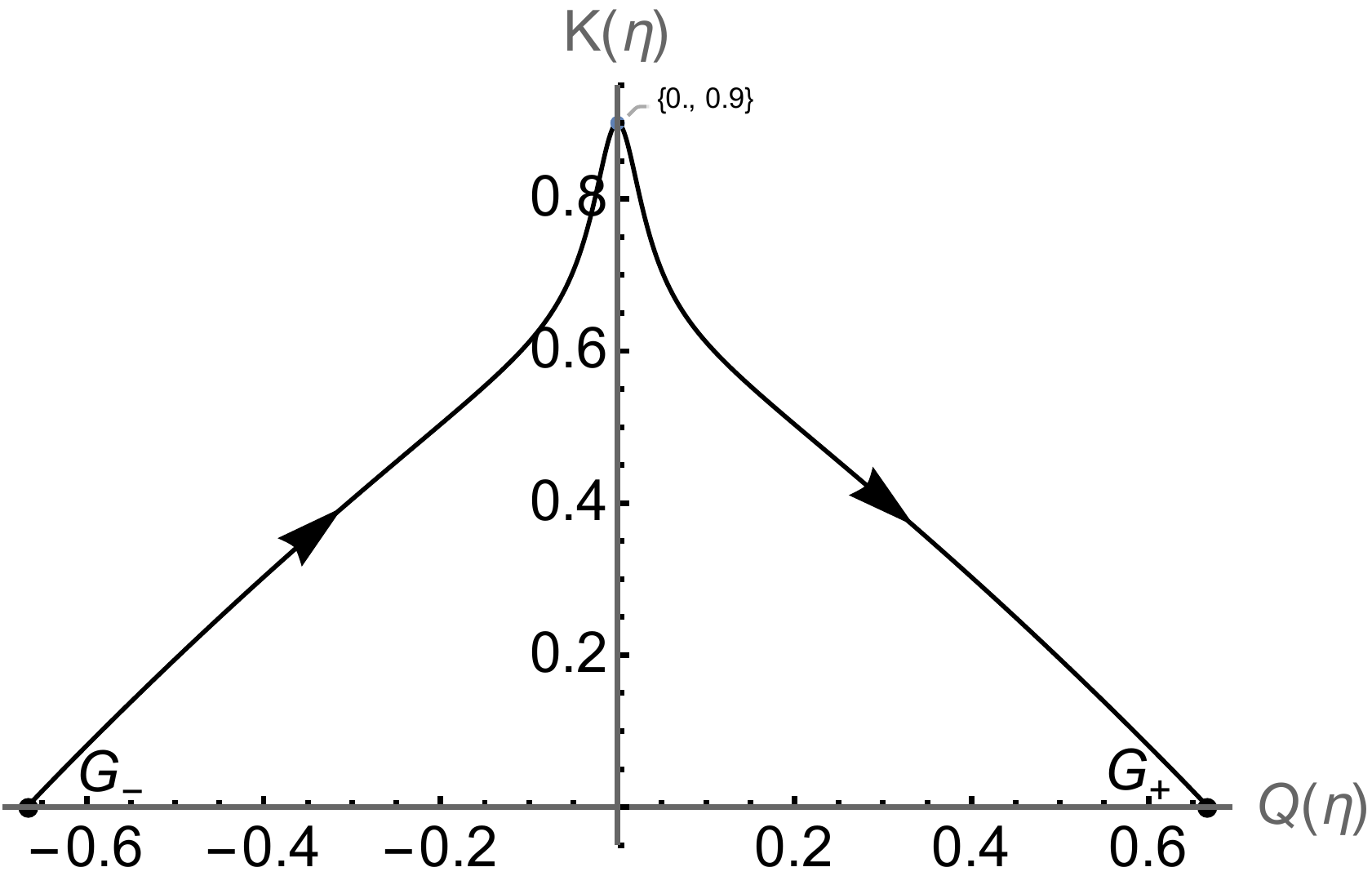}}
    \caption{Parametric plots corresponding to the initial conditions $(K(0),x(0),v(0),Q(0))=(0.9,0,0.95,0)$ (coordinates labelled). The corresponding point in the phase space is marked with a blue dot in Fig.\ref{fig:kneq02Dc}. The trajectory evolves asymptotically from the past attractor $\mathcal{G}_-$ to the future attractor $\mathcal{G}_+$, undergoing a bounce at $(K,x,v,Q)=(0.9,0,0.95,0)$ (since $v>K$ at this point; see Eq.\eqref{bounce_cond}).}
    \label{fig:k09parametric}
\end{figure*}

In Fig.\ref{fig:k1/5parametric}, \ref{fig:k3/5parametric} and \ref{fig:k09parametric} we show explicitly 2-D parametric plots of bouncing and recollapsing trajectories taking the initial conditions from the 2-dimensional slices in Fig. \ref{fig:kneq02D}. 
\begin{figure*}
    \centering
    \subfigure[]{\includegraphics[width=0.3\linewidth]{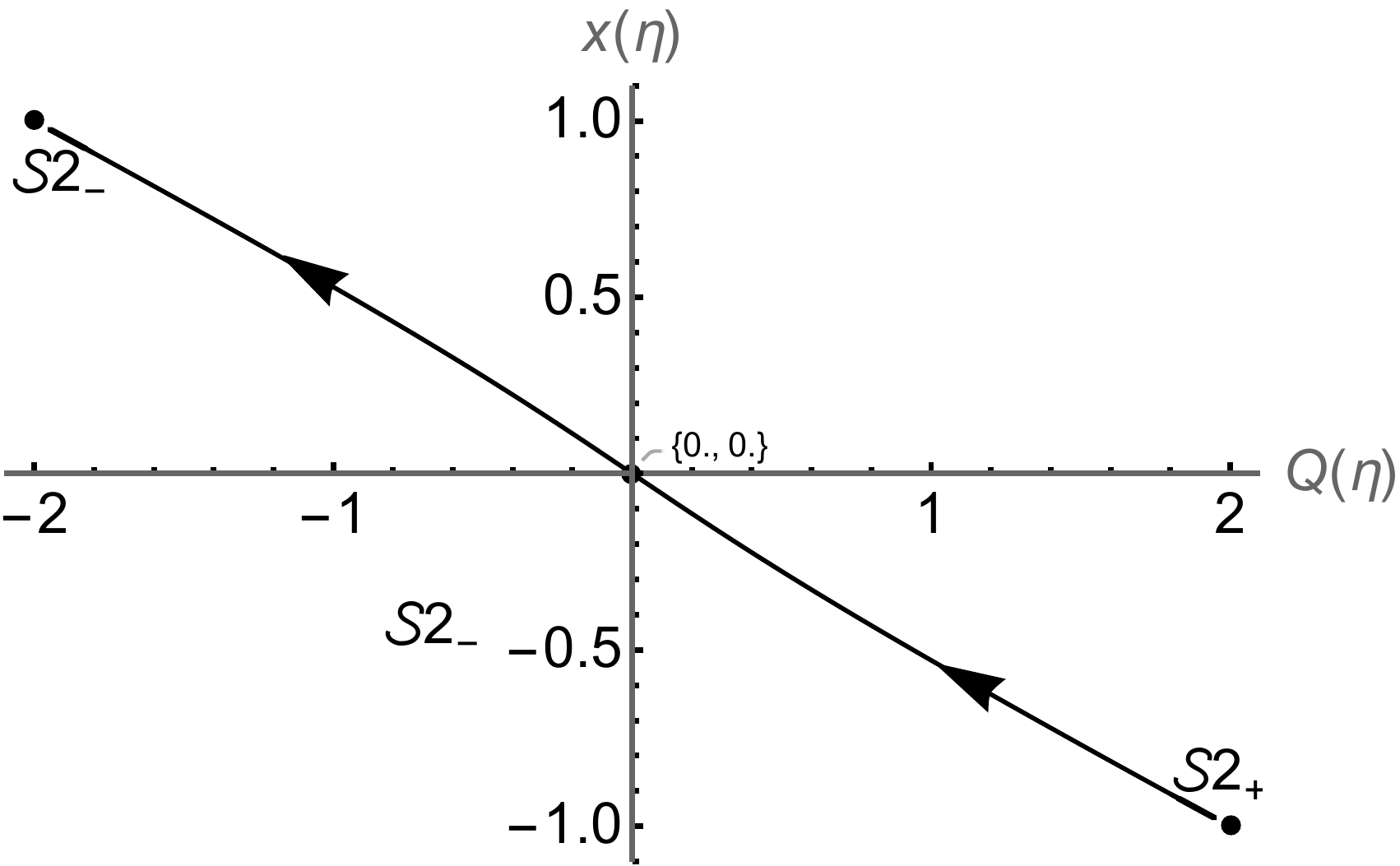}}
    \hspace{0.3cm}
    \subfigure[]{\includegraphics[width=0.3\linewidth]{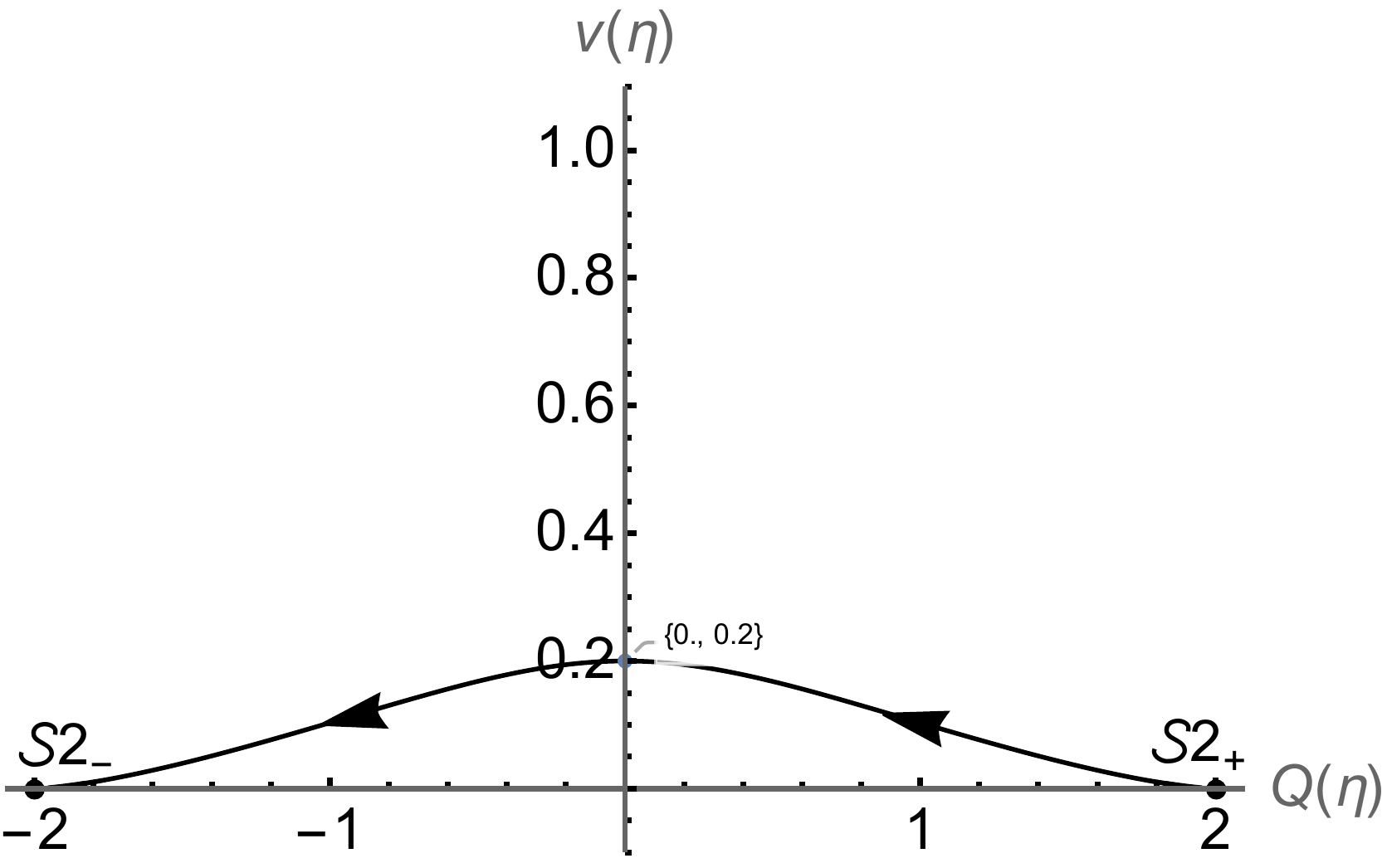}}
    \hspace{0.3cm}
    \subfigure[]{\includegraphics[width=0.3\linewidth]{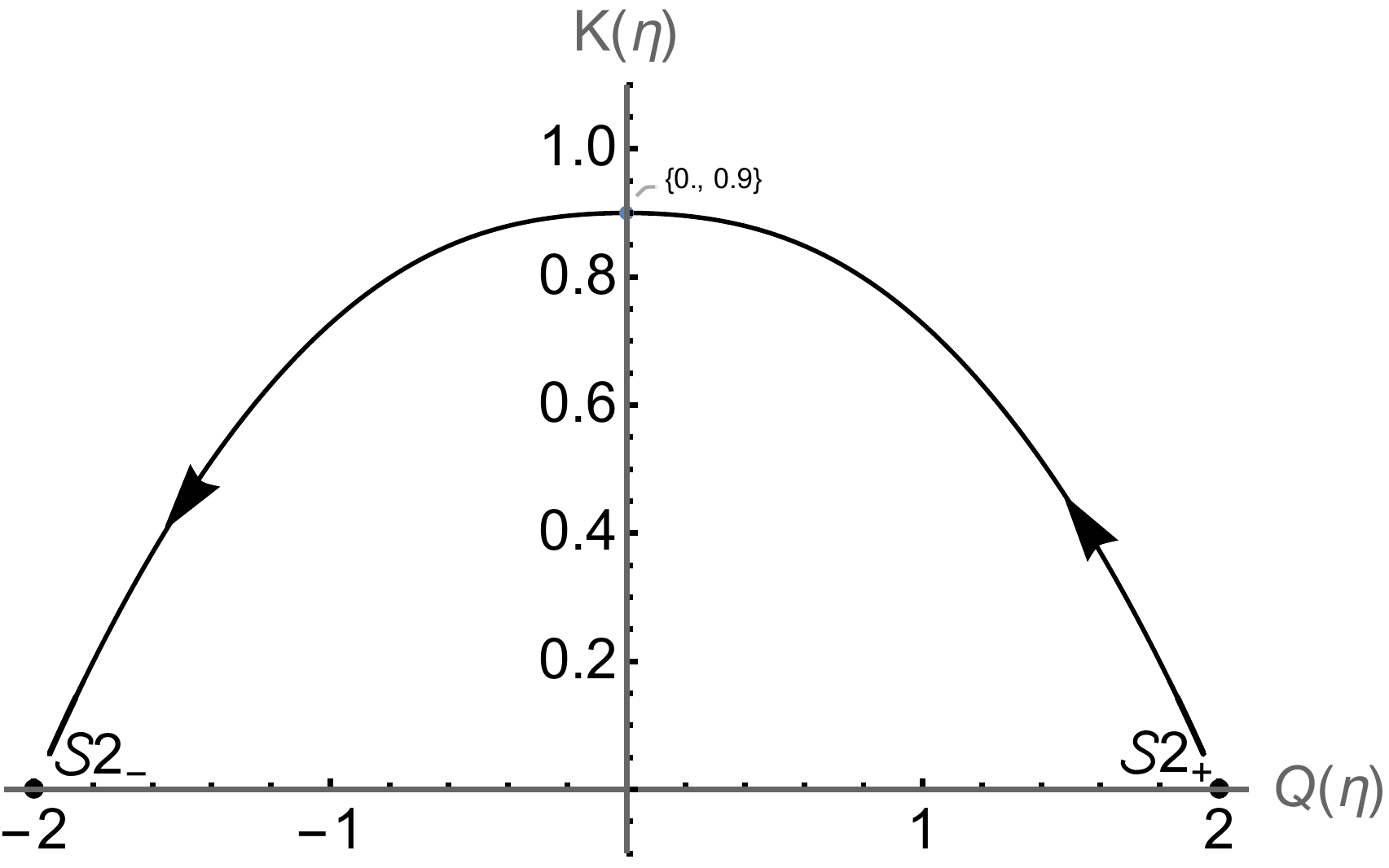}}
    \caption{Parametric plots corresponding to the initial conditions $(K(0),x(0),v(0),Q(0))=(0.9,0,0.2,0)$ (coordinates labelled). The corresponding point is marked with a blue dot in Fig.\ref{fig:kneq02Dc}. The trajectory evolves asymptotically from the saddle fixed point $(K,x,v,Q)=(0,-1,0,2)$ on $\mathcal{S}2_+$ in the past to the saddle fixed point $(K,x,v,Q)=(0,1,0,-2)$ on $\mathcal{S}2_-$ in the future, undergoing a recollapse at $(K,x,v,Q)=(0.9,0,0.2,0)$ (since $v<K$ at this point; see Eq.\eqref{bounce_cond}).}
    \label{fig:k09(2)parametric}
\end{figure*}
Here we find it worth mentioning that cyclic cosmologies are possible in the HS model with $n=1=C_1$. This is the case, for example, for the evolution represented by the parametric plots in Fig.\ref{fig:k1/5parametric}; it periodically undergoes a distinct cycle. Cyclic universes have been studied in the context of $f(R)$-gravity \cite{Pavlovi2021,Ahmed2019ACU,Nojiri2012}, however most results produce numerical solutions or complex analytical functions for the form of $f(R)$ for which such a solution is possible. As far as we are aware, explicit cyclic solutions have not been investigated specifically for physically viable $f(R)$ models like Hu-Sawicki. 

\subsection{$\Lambda$CDM point for the Hu-Sawicki $f(R)$ model}
\label{subsec:lcdm_hs}
\begin{figure*}
    \centering
    \subfigure[]{\includegraphics[width=0.32\linewidth]{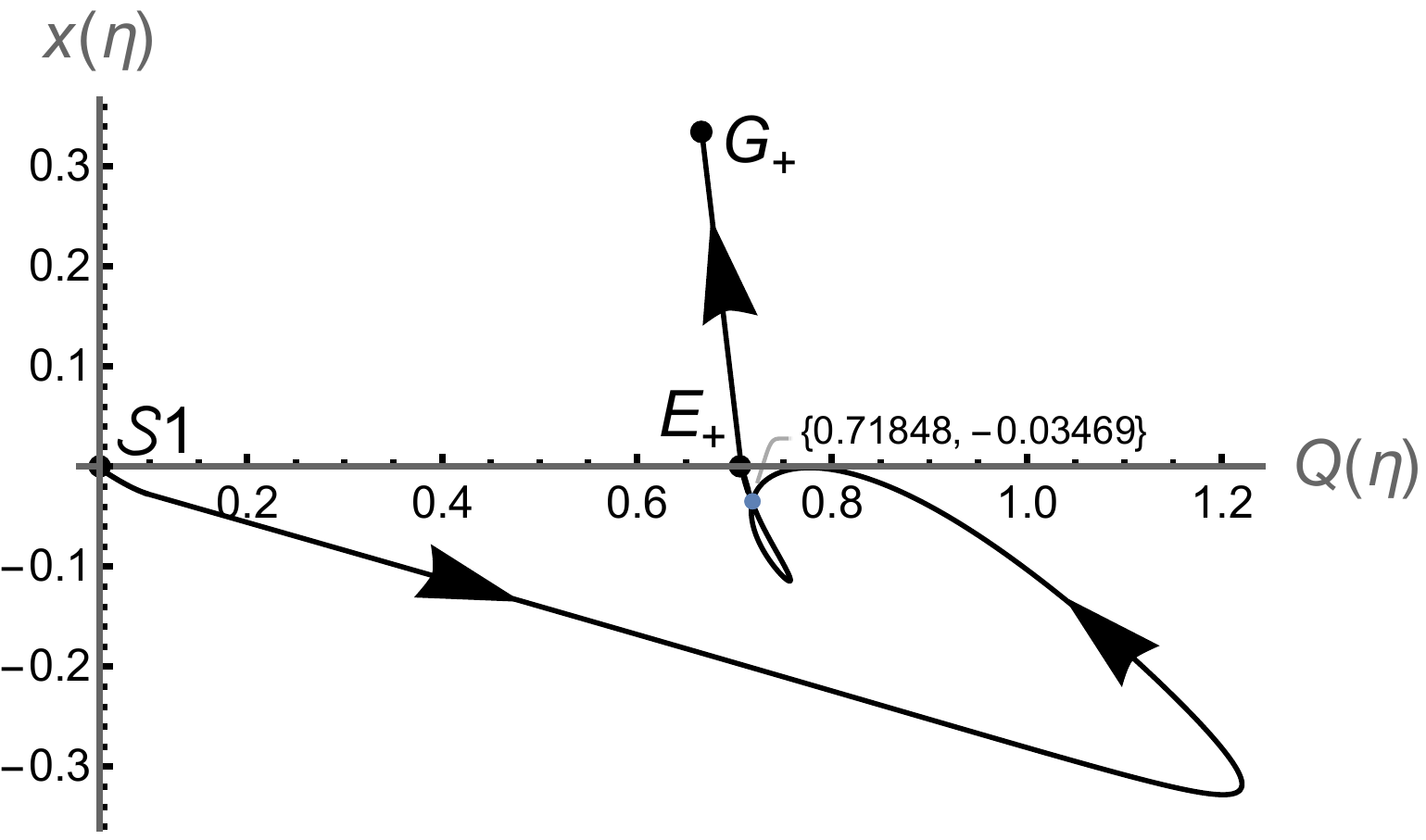}}
    \hspace{0.15cm}
    \subfigure[]{\includegraphics[width=0.32\linewidth]{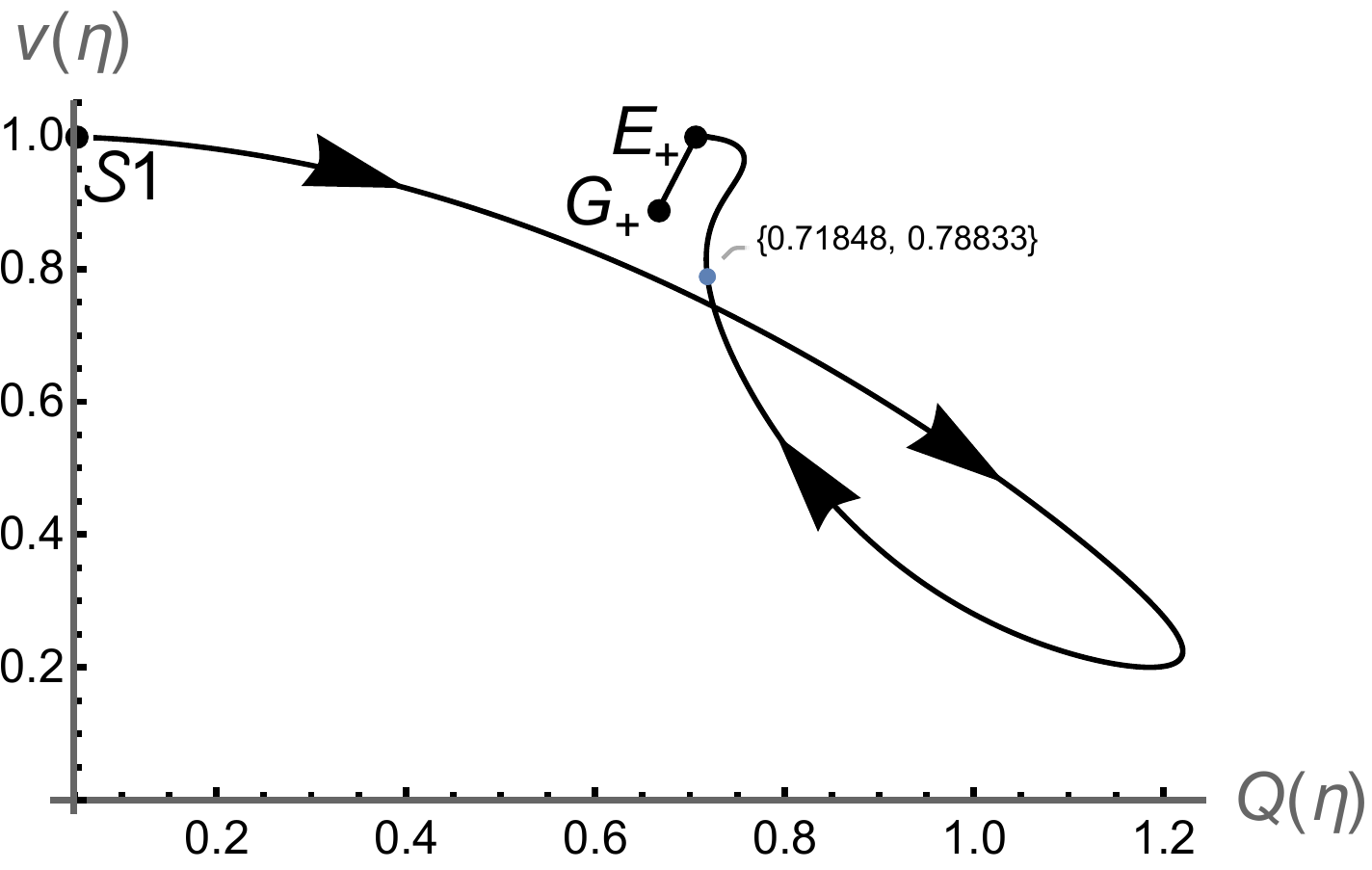}}
    \hspace{0.15cm}
    \subfigure[]{\includegraphics[width=0.32\linewidth]{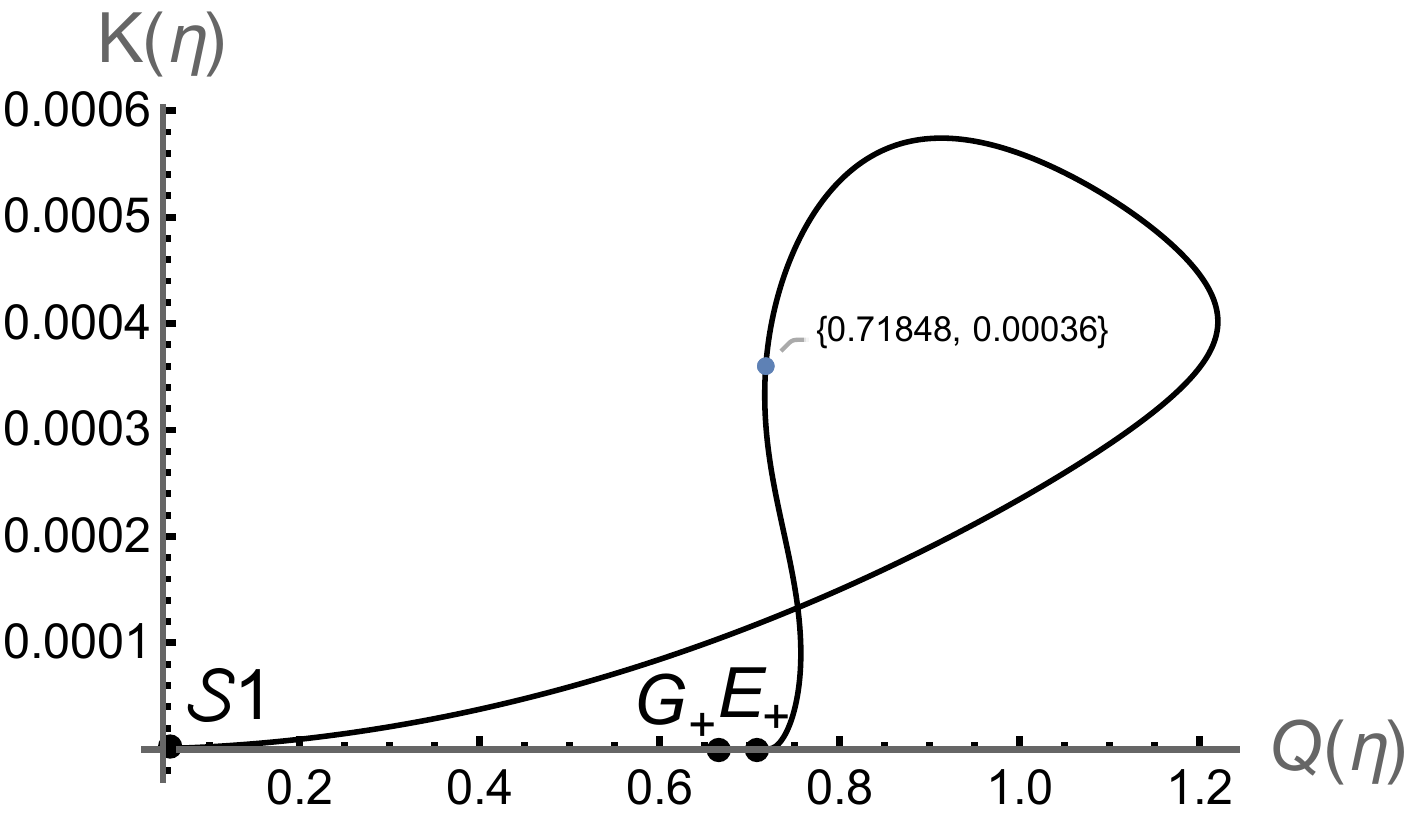}}
    \caption{Parametric plots corresponding to the initial conditions $(K(0),x(0),v(0),Q(0))=(0.00036,-0.03469,0.78833,0.71848)$, which is the $\Lambda$CDM point for the HS model with $n=C_1=1$ (Eq.\eqref{eq: current epoch IC}). The trajectory evolves asymptotically from the point $(K,x,v,Q)=(0,0,1,0)$ on $\mathcal{S}_{1+}$ in the past, passing through the $\Lambda$CDM point and the point $\mathcal{E}_+$ and evolving asymptotically to $\mathcal{G}_+$ in the future.}
    \label{fig:LCDMparametric}
\end{figure*}

The particular case of the Hu-Sawicki model we are considering (\eqref{HS}) has $n=C_1=1$ and only one unspecified model parameter $C_2$. Following the discussion of the last paragraph of Sec.\ref{subsec:lcdm_pt_gen}, one can use Eq.\eqref{eq:initcalc1} to determine $C_2$, which gives a third order equation for $c=C_{2}H_0^2$. However, only one of the three solutions is positive. Since we restrict this study to $C_2>0$, the only viable solution is\footnote{We give the value here to such a degree of accuracy since the dynamical system is very sensitive. It is possible to keep only up to the third decimal place, however this slightly alters some of the values found in subsequent calculations. }
\begin{equation}
   c =  C_2 H_0^2 =0.11841.
\end{equation}
One can then calculate values for $f(R_0)$ and its derivatives (up to second order since this is the highest order derivative required for calculation of the dynamical variable values):
\begin{equation}
\begin{aligned}
    &\frac{f(R_0)}{H_0^2} = 4.76818,\\
    &f'(R_0) = 0.76996,\\
    &\frac{f''(R_0)}{(H_0^2)^{-1}}= 0.02613.
\end{aligned}
\end{equation}
Using these values one can calculate
\begin{equation}
    \begin{aligned}
    D_0 =&\sqrt{\left(3 H_0+\frac{3}{2} \frac{f''(R_0)\dot{R}_0}{f^{\prime}(R_0)}\right)^{2}+\frac{3}{2}\left(\frac{f(R_0)}{f^{\prime}(R_0)}+\frac{6 k}{a_0^{2}}\right)}\\
    =& 4.17549 H_0,
\end{aligned}
\end{equation}
and, therefore, the present day values of the compact dynamical variables
\begin{widetext}
\begin{equation}
\begin{aligned}
(x_0,v_0,y_0,Q_0,K_0,\Omega_0)&=\left(\frac{3}{2} \frac{f''(R_0)\dot{R_0}}{f^{\prime}(R_0)} \frac{1}{D_0},\frac{3}{2} \frac{R_0}{D_0^{2}},\frac{3}{2} \frac{f(R_0)}{f^{\prime}(R_0)} \frac{1}{D_0^{2}},\frac{3 H_0}{D_0},1-(Q_0+x_0)^2-y_0,1-x_0^2-v_0\right)\\
&=(-0.03469,0.78833,0.53207,0.71848,0.00036,0.21047).
\label{eq: current epoch IC}
\end{aligned}
\end{equation}
\end{widetext}
This is the $\Lambda$CDM point for HS model with $n=C_1=1$. In Fig. \ref{fig:LCDMparametric} we show parametric plots taking the initial conditions at this point.
\newpage
\section{Conclusion}
In this paper we have presented a comprehensive analysis of the phase space of $\{n,C_1\}=\{1,1\}$ Hu-Sawicki $f(R)$-gravity model including spatial curvature. We began by extending the compactified dynamical system formalism for a general $f(R)$, presented in \cite{Kandhai:2015pyr}, to include non-spatially flat cosmologies. We then introduced a method to determine a point and surface in the phase space corresponding to the present day universe and the $\Lambda$CDM evolution history respectively (referred to as the $\Lambda$CDM point and surface). Applying these procedures to the Hu-Sawicki model, we identified some novel features of the phase space; including invariant submanifolds, a physically viable region and the existence and stability of two 2-dimensional sheets of fixed points. We presented a trajectory mimicking the $\Lambda$CDM evolution history and compared the evolution of their cosmographic parameters. Following this, we investigated the existence of bouncing and recollapsing cosmologies in the model and arrived at a \emph{no-go} conclusion stating that this kind of trajectory is not possible in the spatially flat model for our specific parameter choice. We also presented the case of a cyclic cosmology which, to the best of our knowledge, has not been identified before in this particular model. 

A comment here is in order regarding Ref.\cite{Amani:2015upn}, in which an analytical method is developed for identifying bouncing solutions in $f(R)$ theory for spatially flat FLRW cosmology. This study finds that the model used (a similar late-time model) does permit bounces for the spatially flat case. The method developed here provides a way to analytically test the \emph{no-go} conclusion for the Hu-Sawicki model with $n=1=C_1$. Applying the analytic method to this model, we find that it is possible to find bounces in the spatially flat case, however this requires that the matter Equation of State parameter $w_m$ cross the phantom barrier ($w_m<-1$). On the surface, it appears that bounces are possible with $-1\leq w_m \leq 1$ but upon further investigation this results in ghost instabilities since $f'<0$. Therefore, if one requires that $w_m$ is non-phantom and that $f'>0$ throughout the evolution history, the \emph{no-go} conclusion holds for the $n=1=C_1$ Hu-Sawicki model.

\section*{Acknowledgements}
KM thanks the University of Cape Town for financial support. This work is based on the research supported in part by the National Research Foundation of South Africa (Grant Number: 123055). S.C. acknowledges the financial assistance provided by the North-West University, South Africa through the postdoctoral grant NWU PDRF Fund NW.1G01487 as well as the accommodation and financial assistance provided kindly by the Department of Mathematics, BITS-Pilani, Hyderabad Campus. PD acknowledges financial support from the First Rand Bank (SA).

\begin{appendix}

\section{Stability of the invariant submanifolds}
\label{app:stabilitysubmanifolds}

Suppose that in the 4-dimensional phase space $K$-$x$-$v$-$Q$, a 3-dimensional invariant submanifold is given by the condition $\mathbb{S}(K,x,v,Q)=0$, i.e.
\begin{equation}
    \frac{d\mathbb{S}}{d\tau}\bigg\vert_{\mathbb{S}=0}=0.
\end{equation}
The stability of this submanifold can be determined as follows. Consider the family of hypersurfaces $\mathbb{S}(K,x,v,Q)=\mathcal{C}$, which contains the invariant submanifold as $\mathcal{C}=0$. Let us consider an arbitrary submanifold with $\mathcal{C}=\delta$ which is infinitesimally close to the invariant submanifold $\mathcal{C}=0$. Keeping in mind the condition for $\mathbb{S}=0$ being an invariant submanifold, Taylor expanding $d\mathbb{S}/d\tau$ around it, one gets
\begin{eqnarray}
  \frac{d\mathbb{S}}{d\tau}\bigg\vert_{\mathbb{S}=\delta} &=& \frac{d(d\mathbb{S}/d\tau)}{d\mathbb{S}}\bigg\vert_{\mathbb{S}=0}\delta + \frac{1}{2}\frac{d^{2}(d\mathbb{S}/d\tau)}{d\mathbb{S}^2}\bigg\vert_{\mathbb{S}=0}\delta^2 + \frac{1}{6}\frac{d^{3}(d\mathbb{S}/d\tau)}{d\mathbb{S}^3}\bigg\vert_{\mathbb{S}=0}\delta^3 +...
\end{eqnarray}
The first nonvanishing term in the above expression gives the directionality (towards or away from) of the flow component normal to the invariant submanifold $\mathbb{S}=0$ for the phase flow in its vicinity, which determines its nature of stability. The invariant submanifold is of attracting (repelling) nature if the phase flow in its vicinity is towards (away from) it.

In the HS model under consideration there are three invariant submanifolds, namely
\begin{equation*}
    K=0, \qquad v=0, \qquad z\equiv v+(Q+x)^{2}+K-1=0.
\end{equation*}

\subsection{Stability of $K=0$}

Using the dynamical equation of $K$, the Taylor expansion of $dK/d\tau$ around $K=0$ up to the first nonvanishing term is 
\begin{widetext}
\begin{equation}\label{stab_flat}
    \frac{dK}{d\tau}\bigg\vert_{K=\delta} = \frac{1}{3}\left[v^{2}x\left((Q+x)^{2}-1\right) - \left((Q+x)^{2}-1+v\right)^{2}\left(-4x+(Q+x)(x^{2}+4xQ+3v-1)\right)\right]\delta + \mathcal{O}(\delta^2)
\end{equation}
\end{widetext}
The invariant submanifold is attracting (repelling) when the coefficient of $\delta$ is negative (positive).
\subsection{Stability of $v=0$}
Using the dynamical equation of $v$, the Taylor expansion of $dv/d\tau$ around $v=0$ up to the first nonvanishing term is 
\begin{widetext}
\begin{equation}
 \frac{dv}{d\tau}\bigg\vert_{v=\delta} = 
  \begin{cases}
     \frac{1}{3}z^{2}\left[Q\left(3-4xQ-5x^{2}\right)+x\left(5-4k-x^{2}\right)\right]\delta
     + \mathcal{O}(\delta^2) \hspace{4em} (\text{For $z>0$}),\\
     \frac{1}{3}x\delta^3 + \mathcal{O}(\delta^4) \hspace{21em}\, (\text{For $z=0$}).
    \end{cases}
\end{equation}
\end{widetext}

The leading nonvanishing term in the Taylor expansion is $\mathcal{O}(\delta)$ for $z>0$ and $\mathcal{O}(\delta^3)$ for $z=0$. The invariant submanifold is attracting (repelling) when the leading coefficient is negative (positive).
\subsection{Stability of $z=0$}
Using Eq.\eqref{dyn_z}, the Taylor expansion of $dz/d\tau$ around $z=0$ up to the first nonvanishing term is 
\begin{equation}
 \frac{dz}{d\tau}\bigg\vert_{z=\delta} = 
  \begin{cases}
     -\frac{1}{3}v^{3}x\delta + \mathcal{O}(\delta^2) \hspace{5em} (\text{For $x\neq0,\,v>0$}), \\
     -\frac{2}{3}Qv\delta^2 + \mathcal{O}(\delta^3) \hspace{4em}\;\,\ (\text{For $x=0,\,v>0$}),\\
     -(2x-6Q)\delta^3 + \mathcal{O}(\delta^4) \hspace{2em}\ (\text{For $v=0$}).
    \end{cases}
\end{equation}
The leading nonvanishing term in the Taylor expansion is $\mathcal{O}(\delta)$ for $x\neq0,v\neq0$, $\mathcal{O}(\delta^2)$ for $x=0,v\neq0$ and $\mathcal{O}(\delta^3)$ for $x\neq0,v=0$.
\begin{itemize}
    \item For $x\neq0,\,v>0$, the invariant submanifold is attracting (repelling) for $x>0$ ($x<0$), with $x=0$ acting as a separatrix.
    \item For $x=0,v>0$, the invariant submanifold $z=0$ has different nature on both sides of it, with $z=0$ serving as a node. Since $C_2>0$ in the HS model under consideration, one recalls from Eq.\eqref{ps_const_1} that the physical viability condition $f''>0$ requires $z>0$. If one confines the attention to the positive side of $z$, then, for $x=0,\,v>0$, the invariant submanifold $z=0$ is attracting (repelling) for $Q>0$ ($Q<0$). 
    \item For $v=0$, the invariant submanifold is attracting (repelling) for $x>3Q$ ($x<3Q$), with $x=3Q$ acting as a separatrix.
\end{itemize}
One can note that the nature of stability of the invariant submanifold $z=0$ is independent of $K$.
\section{Stability of the sheets of fixed points $\mathcal{S}_1,\,\mathcal{S}_2$}
\label{app:stabilitysheets}
From Eqs.\eqref{S1},\eqref{S2} we notice that both the sheets of fixed points $\mathcal{S}_1,\,\mathcal{S}_2$ lie on the invariant submanifold $z=0$. In fact, if we replace $Q$ by $z$ using the definition of $z$
\begin{equation}\label{Qz}
    Q = -x\pm\sqrt{1-v-K+z},
\end{equation}
and use $(K,x,v,z)$ instead of $(K,x,v,Q)$ to be the set of independent dynamical variables, then the equations for the sheets of fixed points are greatly simplified
\begin{eqnarray}
&& \mathcal{S}_1 \equiv (K,x,v,z) = \left(K,0,v,0\right),
\\
&& \mathcal{S}_2 \equiv (K,x,v,z) = \left(K,x,0,0\right).
\end{eqnarray}
In general, there can be phase flows within the invariant submanifold $z=0$ (which represents the limit $f(R)\rightarrow-2\Lambda+R$), but the flow is contained within it. There are, however, parts of $z=0$, given by $\mathcal{S}_1$ and $\mathcal{S}_2$, where there is no flow, as these parts represents sheets of fixed points. In the immediate vicinity of a point $\mathcal{P}\subset\mathcal{S}_1\,\text{or}\,\mathcal{S}_2$, the phase flow is essentially 2-dimensional; If $\mathcal{P}\subset\mathcal{S}_1$, the flow is entirely on the $x$-$z$ plane and if $\mathcal{P}\subset\mathcal{S}_2$, the flow is entirely on the $v$-$z$ plane. This geometrical insight dictates that to investigate the stability of $\mathcal{S}_1,\,\mathcal{S}_2$, one is only required to considered the phase flow in the vicinity of the submanifolds $z=0$, $x=0$, $v=0$, something that we had already considered in Appendix \ref{app:stabilitysubmanifolds}. 
\subsection{Stability of $\mathcal{S}_1$}
The 2-dimensional sheet $\mathcal{S}_1$ lies at the intersection of the 3-dimensional submanifold $x=0$ and the 3-dimensional invariant submanifold $z=0$. In the immediate vicinity of any point on $\mathcal{S}_1$, the phase flow is essentially planar, lying on the $x$-$z$ plane. The nature of stability of the invariant submanifold $z=0$ for $x=0$ has already been discussed in the last subsection of Appendix \ref{app:stabilitysubmanifolds}. In particular, for $v>0$, the flow is towards (away from) $z=0$ for $Q>0$ ($Q<0$). Phase flow in the vicinity of the submanifold $x=0$ for $z=0$ can be found by Taylor expanding $dx/d\eta$ around $x=0$, setting $z=0$ and looking at the leading order nonvanishing term
\begin{equation}
    \frac{dx}{d\eta}\bigg\vert_{x=\delta} = - \frac{1}{6}v^{3}\delta^2 + \mathcal{O}(\delta^3) \qquad (\text{For $z=0,\,v>0$}).
\end{equation}
The flow is towards $x=0$ on the positive side of $x$ and away from $x=0$ on the negative side of $x$, with $x=0$ acting as a node, an inference that holds true for all values of $Q$. Summarizing, one can infer that the sheet of fixed points $\mathcal{S}_1$ is always
\begin{itemize}
    \item attracting on the positive side of $x$.
    \item repelling on the negative side of $x$.
\end{itemize}
Based on this fact alone, one can conclude that all the points on $\mathcal{S}_1$ are saddle fixed points, with there being always a flow component along the $x$-direction from the positive to the negative side of $x$ past $x=0$ in the vicinity of any point $\mathcal{P}\subset\mathcal{S}_1$. One can note that the stability of $\mathcal{S}_1$ is completely independent of $K$ and $v$, i.e. the same nature of stability holds everywhere on $\mathcal{S}_1$.
\subsection{Stability of $\mathbb{S}_2$}
The 2-dimensional sheet $\mathcal{S}_2$ lies at the intersection of two 3-dimensional invariant submanifolds $v=0$ and $z=0$. In the immediate vicinity of any point on $\mathcal{S}_2$, the phase flow is essentially planar, lying on the $v$-$z$ plane. The stability of both $v=0$ and $z=0$ have already been discussed in Appendix \ref{app:stabilitysubmanifolds}. For $z=0$, the invariant submanifold $v=0$ is attracting (repelling) for $x<0$ ($x>0$). For $v=0$, the invariant submanifold $z=0$ is attracting (repelling) for $x>3Q$ ($x<3Q$). Summarizing, one can infer that the sheet of fixed points $\mathcal{S}_2$ is 
\begin{itemize}
    \item attracting for $3Q<x<0$.
    \item repelling for $0<x<3Q$.
    \item saddle otherwise.
\end{itemize}
Given any point on $\mathcal{S}_2$, one can now infer about it's stability. Note that the stability of $\mathcal{S}_2$ depends on $x$ as well as $K$ (by virtue of Eq.\eqref{Qz}).
\end{appendix}
\section*{References}
\bibliographystyle{unsrt}
\bibliography{refs}

\end{document}